\documentclass[conference]{IEEEtran}
\usepackage{tugraz_defaults}
\usepackage{enumitem}
\newcommand{\parhead}[1]{\noindent\textbf{#1.}\ }
\begin{document}

\title{
  Another Flip in the Wall of Rowhammer Defenses
}
\author{
\IEEEauthorblockN{Daniel Gruss$^1$, Moritz Lipp$^1$, Michael Schwarz$^1$, Daniel Genkin$^2$,\\Jonas Juffinger$^1$, Sioli O'Connell$^3$, Wolfgang Schoechl$^1$, and Yuval Yarom$^{3,4}$}
\IEEEauthorblockA{$^1$ Graz University of Technology \\ $^2$ University of Pennsylvania and University of Maryland \\ $^3$ University of Adelaide \\ $^4$ Data61  \vspace{-0.2cm}}
}
\newcommand{\citeA}[1]{\citeauthor{#1}~\cite{#1}}
\newcommand{\daniel}[1]{\dtcolornote[Daniel]{blue}{#1}}
\maketitle

\begin{abstract}
The Rowhammer bug allows unauthorized modification of bits in DRAM cells from unprivileged software, enabling powerful privilege-escalation attacks.
Sophisticated Rowhammer countermeasures have been presented, aiming at mitigating the Rowhammer bug or its exploitation.
However, the state of the art provides insufficient insight on the completeness of these defenses.

In this paper, we present novel Rowhammer attack and exploitation primitives, showing that even a combination of all defenses is ineffective.
Our new attack technique, \emph{one-location hammering}, breaks previous assumptions on requirements for triggering the Rowhammer bug, \ie we do not hammer multiple DRAM rows but only keep one DRAM row constantly open.
Our new exploitation technique, \emph{opcode flipping}, bypasses recent isolation mechanisms by flipping bits in a predictable and targeted way in userspace binaries.
We replace conspicuous and memory-exhausting spraying and grooming techniques with a novel reliable technique called \emph{memory waylaying}.
Memory waylaying exploits system-level optimizations and a side channel to coax the operating system into placing target pages at attacker-chosen physical locations.
Finally, we abuse Intel SGX to hide the attack entirely from the user and the operating system, making any inspection or detection of the attack infeasible.
Our Rowhammer enclave can be used for coordinated denial-of-service attacks in the cloud and for privilege escalation on personal computers.
We demonstrate that our attacks evade all previously proposed countermeasures for commodity systems.
\end{abstract}

\bgroup
\let\thefootnote\relax\footnotetext{Preprint of the work accepted at the 39th IEEE Symposium on Security and Privacy 2018 (\url{https://www.ieee-security.org/TC/SP2018/index.html}).}
\egroup

\section{Introduction}\label{sec:intro}
The Rowhammer bug is a hardware reliability issue in which an attacker repeatedly accesses (\emph{hammers}) DRAM cells to cause unauthorized changes in physically adjacent memory locations.
Since its initial discovery as a security issue~\cite{Kim2014}, Rowhammer's ability to defy abstraction barriers between different security domains  has been extensively used for mounting devastating attacks on various systems. 
Examples of previous attacks include privilege escalation, from native environments~\cite{Seaborn2015BH}, from within a browser's sandbox~\cite{Gruss2016Row}, and from within virtual machines running on third-party compute clouds~\cite{Xiao2016}, mounting fault attacks on cryptographic primitives~\cite{Razavi2016,Bhattacharya2016}, and obtaining root privileges on mobile phones~\cite{VanDerVeen2016}. 
Recognizing the apparent danger, these attacks have sparked interest 
in developing effective and efficient mitigation techniques.
While existing hardware countermeasures such as using memory with error-correction codes (ECC-RAM) appear to make Rowhammer attacks harder~\cite{Kim2014}, ECC-RAM is intended for server computers and is typically not supported on consumer-grade machines.

Software-based mitigations, which can be implemented on commodity systems, have also been proposed. 
These include ad-hoc defense techniques such as doubling the RAM refresh rates~\cite{Kim2014}, removing unprivileged access to the \texttt{pagemap} interface~\cite{ShutemovLinux2015,Seaborn2015BH,Salyzyn2015}, and prohibiting the \texttt{clflush} instruction~\cite{Seaborn2015BH}. 
However, recent works have already bypassed these countermeasures~\cite{Aweke2016,VanDerVeen2016,Gruss2016Row}. 
Other ad-hoc attempts, such as disabling page deduplication by default~\cite{WindowsServerDeduplication, RedHatKSM}, only prevent specific Rowhammer attacks exploiting these features~\cite{Bosman2016,Razavi2016}, but not all Rowhammer attacks.

The research community proposed sophisticated defenses which seemingly have solved the Rowhammer problem.
Based on the underlying primitives of these defenses, we introduce a new systematic categorization into five defense classes:
\begin{itemize}[nolistsep,align=left, leftmargin=9pt, labelwidth=0pt, itemindent=!]
\item \parhead{Static Analysis} Binary code is analyzed for specific behavior, common in side-channel attacks, \eg using high-resolution timers or cache flush instructions~\cite{Irazoqui2016mascat,Herath2015}.
\item \parhead{Monitoring Performance Counters}
Rowhammer relies on frequent accesses to DRAM cells, \eg using a \FlushReload loop.
These frequent accesses are detected by monitoring CPU performance counters~\cite{Aweke2016,Irazoqui2016mascat,Herath2015,Gruss2016Flush,Payer2016,Chiappetta2015,Zhang2016CloudRadar}.
\item \parhead{Monitoring Memory Access Patterns} Rowhammer causes unusual high-frequency memory access patterns to two or more addresses in one DRAM bank.
Rowhammer can be stopped by detecting such access patterns~\cite{Aweke2016,Corbet2016}.
\item \parhead{Preventing Exhaustion-based Page Placement} Rowhammer requires target pages to be on vulnerable memory locations.
All Rowhammer privilege escalation attacks so far required memory exhaustion.
Thus, preventing abuse of memory exhaustion thwarts Rowhammer attacks~\cite{VanDerVeen2016,Gruss2016Row}.
\item \parhead{Preventing Physical Proximity to Kernel Pages} As a more complete solution, user and kernel memory cells are physically isolated through the memory allocator, thwarting all practical Rowhammer privilege-escalation attacks~\cite{Brasser2017catt}.
\end{itemize}
Notice that defenses in each class share the same assumptions, properties, and introduce the same form of protection. 
Defenses from different classes complement each other. 
Thus, given the extensive amount of research on Rowhammer countermeasures, in this paper we ask the following question:

{\centering{\emph{
To what extent do the approaches above actually prevent Rowhammer attacks? 
In particular, is it possible to successfully mount Rowhammer privilege-escalation attacks in the presence of some (or even all) of the countermeasures above?
}
}}

\subsection{Our Results and Contributions}
In this paper, we show that despite numerous works on mitigating Rowhammer attacks, much remains to be done to truly understand their effectiveness and how to mitigate them. 
For this purpose, we introduce a new categorization for Rowhammer defenses (which we already outlined above) as a foundation for a systematic evaluation.
Demonstrating the insufficiency of existing mitigation techniques, we present a novel Rowhammer attack and subsequent exploitation techniques for privilege escalation which allows defeating the underlying assumptions of all of the countermeasures mentioned above. 
In particular, our attack is still applicable even in the presence of \emph{all} of the above countermeasures. 
We now describe the four building blocks of our attack and how each building block invalidates the assumptions of the defense classes. 

\parhead{Defeating Physical Kernel Isolation}
The assumption of physical kernel isolation is that Rowhammer-based privilege escalation is only practical by flipping bits in kernel pages.
We void this assumption by introducing \emph{opcode flipping}, a technique for malicious and unauthorized modification of a userspace program's instructions by causing bit flips in its opcodes. 
By applying this technique to \texttt{sudo}, we bypass authentication checks and obtain root privileges. 

\parhead{Defeating Memory Access Pattern Analysis}
All known Rowhammer techniques require frequent alternating accesses to \emph{two} or more DRAM cells in the same DRAM bank.
Consequently, countermeasures detect when an attacker performs such alternating accesses to \emph{two} or more addresses in the same DRAM bank. 
We present \emph{one-location hammering}, a new type of Rowhammer attack which only hammers \emph{one} single address.
Since our attack only uses \emph{one} memory address, it does not require any knowledge of physical addresses and DRAM mappings~\cite{Pessl2016,Xiao2016,Jung2016}, allowing us to perform Rowhammer attacks with even fewer requirements.

\parhead{Page Placement Without Memory Exhaustion}
Page deduplication is usually disabled for security reasons~\cite{WindowsServerDeduplication, RedHatKSM, VanDerVeen2016} as a response to page deduplication attacks~\cite{Barresi2015, Suzaki2011, Gruss2015dedup}, including deduplication-based Rowhammer attacks~\cite{Bosman2016,Razavi2016}, 
Hence, attacks can only use memory exhaustion~\cite{Seaborn2015BH,Xiao2016,Gruss2016Row,VanDerVeen2016} to surgically place a target page on a vulnerable physical memory location.
Consequently, countermeasures aim to prevent adversarial memory exhaustion~\cite{VanDerVeen2016,Gruss2016Row}.
We introduce \emph{memory waylaying}, a reliable technique exploiting the Operating System (OS) page cache to influence the physical location of a target page. 
Unlike previous techniques, memory waylaying does not exhaust the system memory and does not cause out-of-memory situations, \ie the system remains stable and responsive.

\parhead{Defeating Countermeasures based on Performance Counters and Static Analysis}
SGX is an x86 instruction-set extension to securely and confidentially run programs in isolated environments, called \emph{enclaves}, on potentially adversary-controlled systems.
Enclaves run with regular user privileges and are further restricted for their own security and safety, \eg no system calls.
To protect against compromised or malicious OSs and hardware, the memory of the enclave is encrypted to prevent any modification or inspection of the enclave's memory contents, even by the OS's kernel and hardware components~\cite{Costan2016}. 
Furthermore, enclaves are excluded from the CPU performance counters~\cite{Schwarz2017SGX}.
Hence, this approach defeats countermeasures which rely on monitoring performance counters~\cite{Aweke2016,Herath2015,Gruss2016Flush,Payer2016} or on analyzing the application code or instruction stream for Rowhammer attacks~\cite{Irazoqui2016mascat,Herath2015}.  

\subsection{Attack Scenarios} 
Our attacks apply to personal computers and cloud systems.
Hence, we demonstrate our attacks in both of these scenarios. 
\begin{itemize}[nolistsep,align=left, leftmargin=9pt, labelwidth=0pt, itemindent=!]
	\item
\parhead{Native Privilege Escalation Attack}
Our Rowhammer enclave can be used on personal computers to gain root privileges on the system, even in the presence of \emph{all} of the defenses mentioned above.
\item \parhead{A Cloud Denial-of-Service Attack}
Our Rowhammer enclave can also be used in the cloud, to shut down a large number of cloud machines in a coordinated way, \ie a ``distributed'' denial-of-service attack, by abusing Intel SGX security mechanisms. 
When SGX detects an error in the encrypted and integrity-checked memory region, it halts the entire machine until a manual power cycle is performed. 
By coordinating the error injection over multiple machines, an attacker can potentially take down an entire cloud provider.
\end{itemize}

\subsection{Paper Outline}
\cref{sec:background} provides background.
\cref{sec:categorization} introduces a new categorization of Rowhammer defenses.
\cref{sec:threat_model} defines our attacker model.
\cref{sec:highlevel} overviews our attack and its building blocks, which are detailed in \cref{sec:opcode} (opcode flipping), \cref{sec:zero} (one-location hammering), and \cref{sec:waylaying} (memory waylaying).
\cref{sec:attacks} evaluates our attacks in practical scenarios.
\cref{sec:discussion} discusses limitations and additional observations.
We conclude in \cref{sec:conclusion}.

\section{Background}\label{sec:background}
In this section, we overview the Rowhammer bug and defenses, discuss the prefetch side-channel attack which we use in \cref{sec:waylaying}, and provide background on Intel SGX.

\subsection{The Rowhammer Bug}\label{sec:bg_rh_att}
The increase in density and decrease in size of DRAM cells leads to smaller capacitance of cells, allowing them to operate using lower voltages and smaller charges. 
While these changes have many advantages, such as an increase in DRAM capacity and lower energy consumption, they also cause DRAM cells to become more susceptible to disturbance errors and unintended physical interactions between multiple cells. 
Such interactions and disturbances often cause memory corruption, where the bit-value of a DRAM cell is unintentionally flipped~\cite{Mutlu2017rowhammer}.  

In \citeyear{Kim2014}, \citeA{Kim2014} showed that such bit errors can be caused in a DRAM row by rapidly accessing memory locations in adjacent DRAM rows (also known as \emph{row hammering}~\cite{Huang2012}).
To achieve these rapid DRAM accesses, data-caching mechanisms need to be bypassed, either by flushing the cache, \eg using \texttt{clflush}~\cite{Kim2014}, cache eviction~\cite{Gruss2016Row,Aweke2016,Aga2017}, or uncached memory accesses~\cite{Qiao2016}.
We now describe different Rowhammer techniques to obtain bit flips in the target row.

\emph{Single-sided} hammering performs frequent memory accesses (hammering) to only one row which is adjacent to the target row. 
In contrast, \emph{double-sided} hammering hammers two memory rows, one on each side of the target row. 
As the two hammered rows must be on different sides of the target row, double-sided hammering generally requires at least partial knowledge of virtual-to-physical mappings while single-sided hammering does not.
Both hammering techniques produce abnormal memory access patterns as they induce an enormous number of row conflicts.
Bit flips are highly reproducible: Hammering the same offsets again yields the same bit flips.

Although the name single-sided hammering may suggest that only a single memory location is hammered, \citeA{Seaborn2015BH}, who introduced this technique, hammer 8 memory locations simultaneously.
On their systems, two or more randomly selected addresses (\ie no knowledge of virtual-to-physical mappings is required) are in the same DRAM bank in \SI{61.4}{\percent} of the cases.
Hence, in fact, single-sided hammering aims to hammer two memory locations in the same bank, but not necessarily neighboring the victim row.

Not a privilege-escalation attack but an escape from the NaCl sandbox was demonstrated by \citeA{Seaborn2015BH}.
NaCl executes arbitrary generated code directly on the CPU but sanitizes it using a blacklist, \eg no system calls.
To bypass the sanitizer, the attacker generates and sprays unprivileged code over the entire memory and induces an unpredictable random bit flip at an unpredictable random memory location.
With a low probability, the bit flip hits the operand of an \texttt{and} instruction used to sanitize addresses used by the sandboxed code.
As the code can be read and executed by the attacker, the attacker can verify whether the random bit flip modified a random code location such pointers are not fully sanitized, re-enabling traditional control-flow diversion attacks.
\citeA{Bhattacharya2016} exploited random Rowhammer bit flips in random memory locations to produce faulty RSA signatures, to recover the secret key.

However, as bit flips are highly reliable, more deterministic and reliable attacks have been mounted, including privilege-escalation attacks, sandbox escapes, and compromise of cryptographic keys were demonstrated using memory spraying~\cite{Seaborn2015BH,Gruss2016Row,Xiao2016}, grooming~\cite{VanDerVeen2016}, or page deduplication~\cite{Bosman2016,Razavi2016}.

\subsection{Rowhammer Defenses}\label{sec:bg_rh_def}
Rowhammer defenses can be divided into three categories based on their goal.
The first category aims to \emph{detect} Rowhammer and, after detection, stop the corresponding processes.
The second category aims to \emph{neutralize} Rowhammer bit flips to prevent their exploitation.
The third category aims to \emph{eliminate} Rowhammer bugs.
We now review previous works on defending against Rowhammer attacks.
We group the proposed countermeasures using the above-mentioned three categories.

\parhead{Rowhammer Detection Countermeasures} Static code a\-na\-ly\-sis could be used to {detect} microarchitectural attacks in binaries in a fully automated way, \eg when tested before loading them into an app store~\cite{Irazoqui2016mascat}.
Several works detect on-going attacks using hardware- and software-based performance counters~\cite{Herath2015, Payer2016, Gruss2016Flush, Corbet2016, Chiappetta2015,Zhang2016CloudRadar}.
\citeA{Herath2015} detect attacks by monitoring suspicious cache activity of processes using performance counters and then searching for \texttt{clflush} instructions near the instruction pointer.

\parhead{Rowhammer Neutralization Countermeasures}
The system's memory allocator only places kernel pages near userspace pages in near-out-of-memory situations.
Hence, modifying the allocator to prefer the out-of-memory situation over the proximate placement of kernel and userspace pages, effectively prevents memory exhaustion in turn of spraying and grooming~\cite{VanDerVeen2016,Gruss2016Row}.
This prevents known Rowhammer attacks based on memory grooming or memory spraying, as the target page cannot be evicted or placed anymore, \ie {neutralizes} Rowhammer bit flips.
Generalizing this, \citeA{Brasser2017catt} presents G-CATT, an alternative memory allocator that isolates user and kernelspace in physical memory ensuring that the attacker cannot exploit bit flips in kernel memory, thus {neutralizing} Rowhammer-induced bit flips.
Disabling page deduplication prevents Rowhammer attacks exploiting these features~\cite{WindowsServerDeduplication, RedHatKSM,Bosman2016,Razavi2016}.

\parhead{Rowhammer Elimination Countermeasures}
ANVIL~\cite{Aweke2016} uses performance counters to detect and subsequently mitigate Rowhammer attacks.  More specifically, ANVIL uses the CPU's  performance counters in order to continuously monitor the amount of cache misses.
When the amount of cache misses exceeds a predetermined threshold, ANVIL's  second stage is initiated, logging the addresses of cache misses.
Finally, ANVIL mitigates Rowhammer effects by selectively refreshing nearby memory rows.
However, as refreshing a row imposes some performance penalties, ANVIL avoids having a large number of false positives by discarding all logged cases that do have a significant amount of accesses to at least two rows in the same memory bank.
While this optimization improves ANVIL's performance, as we discuss in Section~\ref{sec:categorization}, it also prevents ANVIL from detecting one-location hammering, thus facilitating our attack.
Similarly to ANVIL's detection approach, \citeA{Corbet2016} discusses halting the CPU when cache-miss rates exceed a threshold, slowing down not only Rowhammer attacks but the entire system. 

\citeA{Brasser2017catt} also presented B-CATT, a bootloader extension blacklisting vulnerable locations, thus, effectively reducing the amount of usable memory, but fully eliminating the Rowhammer bug.
However, \citeA{Kim2014} observed that this approach is not practical as it would block almost the entire memory.
We validated this observation and found more than \SI{95}{\percent} of the memory would be blocked, on several of our systems.
Eliminating Rowhammer by blacklisting the \texttt{clflush} instruction~\cite{Seaborn2015BH} was shown ineffective with cache-eviction-based Rowhammer attacks~\cite{Gruss2016Row,Aweke2016,Aga2017}.

Besides building more reliable chips or employing ECC modules, \citeA{Kim2014} and \citeA{Kim2015} proposed probabilistic methods to eliminate bit flips in hardware.
Every time a row is opened and closed, other adjacent or non-adjacent rows are opened with a low probability.
Thus, if a Rowhammer attack opens and closes rows, statistically the adjacent rows are refreshed as well and, thus, bit flips are averted.
The LPDDR4 standard~\cite{LPDDR4} specifies two features to {eliminate} the Rowhammer bug: Target Row Refresh (TRR) enables the memory controller to refresh rows adjacent to a certain row; Maximum Activation Count (MAC) specifies how often a row can be activated before adjacent rows need to be refreshed.
Furthermore, \citeA{Ghasempour2015} presented ARMOR, a cache storing frequently accessed rows in order to reduce the number of row activations in the DRAM and, thus, {eliminating} the Rowhammer bug.

Hence, all elimination-based defenses are either not practical or require hardware changes, making them not applicable for commodity systems.
Commodity systems should instead be protected using detection- or neutralization-based approaches.

\subsection{The Prefetch Side-Channel Attack}
The prefetch side-channel attack was presented by \citeA{Gruss2016CCS} as a way to defeat address-space-layout randomization.
The timing difference induced by the prefetch instruction depends on the state of various caches.
Prefetch instructions ignore privileges and permissions. 
Prefetch side-channel attacks also exploit the OS design.
In most OSs, every valid memory location in a user process is mapped at least twice, once in the user process virtual memory, and once in the direct-physical mapping in the kernelspace.
The \emph{prefetch address-translation oracle} exploits this direct-physical mapping to determine whether an address in userspace maps to a specific address in the direct-physical mapping.
If the guess was correct, the attacker learns the physical address of a userspace virtual address.
Hence, the attacker does not have to rely on OS interfaces to obtain physical addresses for virtual addresses.

\subsection{Intel SGX}\label{sec:sgx}

Intel SGX is an x86 instruction-set extension for integrity and confidentiality of code and data in untrusted environments~\cite{Costan2016}.
For this purpose, SGX executes programs in so-called \emph{secure enclaves} which use protected areas of memory that can only be accessed by the enclaves themselves.
With SGX implemented in the CPU, the enclave remains protected, even if OS, hypervisor, and hardware have been compromised.
Furthermore, remote attestation allows validating the integrity of the enclave by proving its correct loading.

Intel SGX explicitly protects against DRAM-based attacks, \eg cold-boot attacks, memory bus snooping, and memory-tampering attacks, by cryptographically ensuring confidentiality, integrity, and freshness of data stored in the main memory.
Hence, it removes the DRAM from the trusted computing base.
The memory containing code and data of running enclaves is a physically contiguous and encrypted block in the DRAM, called \emph{EPC} (enclave page cache) area, which is protected from all non-enclave memory accesses using protection mechanisms implemented in the CPU.
The encryption by the Memory Encryption Engine (MEE) is transparent to the processor's cores~\cite{Gueron2016}.
The MEE utilizes a Merkle tree to detect when the encrypted code and data stored in the DRAM have been tampered with.
The MEE provides freshness to the integrity tags to mitigate replay attacks, \ie replacing a new encrypted page with an old encrypted page.

If an integrity or freshness error occurred, Intel SGX aborts the execution of the memory fetch immediately, and the MEE emits an error signal.
Thus, the unverified data of the DRAM will never be loaded into the last-level cache~\cite{Gueron2016}.
Moreover, the MEE locks the memory controller, preventing any future memory operations (potentially incurring data corruption), causing the system to halt until it is rebooted. 

\subsection{Attacks on (and from) Secure Enclaves}\label{sec:sgxattacks}

While Intel does not claim to protect against side-channel attacks that deduce information of collected power statistics, performance statistics, branch statistics, or information on pages accessed via page tables~\cite{sgxtutorial}, several such attacks have been demonstrated. 
\citeA{Xu2015controlled} demonstrated a page fault side-channel attack from a malicious OS to extract sensitive information, \eg text documents and images.
\citeA{Brasser2017sgx} demonstrated a \PrimeProbe cache side-channel attack, extracting \SI{70}{\percent} of an RSA private key in an enclave.
Furthermore, \citeA{Schwarz2017SGX} mounted a cache side-channel attack from within an enclave to extract a full RSA private key of a co-located enclave.
\citeA{Xiao2017} mounted control-flow inference attacks on recent SSL libraries running in secure enclaves.
\citeA{Moghimi2017} presented CacheZoom, a tool that provides a high-resolution channel to track all memory accesses of SGX enclaves to mount key recovery attacks.
\citeA{Wang2017SGX} systematically analyzed side-channel threats of SGX and identified 8 potential side-channel attack vectors.
However, Intel considers all of these attacks out of scope, due to their side-channel nature.

Attacks that rely on shared memory (\eg \FlushReload~\cite{Yarom2014}) cannot be mounted, as enclave memory is inaccessible for other enclaves, processes, and the OS.
But as DRAM rows are shared, \citeA{Wang2017SGX} showed a cross-enclave DRAMA attack (\cf \cite{Pessl2016}) on other enclaves.

In a concurrent and independent work, Jang~\etal\cite{Jang2017SGXBomb} propose a denial-of-service attack running Rowhammer in an SGX enclave.
We compare their and our observations in \cref{sec:dos}, where we describe a very similar attack.

\section{Categorization of State-of-the-art Defenses for Commodity Systems} \label{sec:categorization}
Discussing Rowhammer defenses based on their goal (detection, neutralization, and elimination; \cf \cref{sec:bg_rh_def}), does not allow a thorough analysis and comparison, as the primitives of the different defenses in each category vary widely.
As we have seen in \cref{sec:bg_rh_def}, none of the elimination-based defenses are practical or applicable to commodity systems.
Hence, in this paper, we only focus on detection- and neutralization-based defenses.
In this section, we introduce a novel systematic categorization for state-of-the-art defenses for commodity systems.

In our evaluation of defenses we identified the following 5 defense classes which can be applied to commodity systems:
\begin{compactenum}
\item[\textbf{D1.}] Detection through \emph{static analysis}.
\item[\textbf{D2.}] Detection through \emph{performance counter analysis}. 
\item[\textbf{D3.}] Detection through analysis of \emph{memory access patterns}.
\item[\textbf{D4.}] Prevention by strictly avoiding \emph{physical proximity}. 
\item[\textbf{D5.}] Prevention by preventing conspicuous \emph{memory footprints}. 
\end{compactenum}
\noindent Other defense classes (bootloader- or BIOS-update-based) have already been shown to be ineffective (\cf \cref{sec:bg_rh_def}), or cannot be applied to commodity systems (hardware modifications).
\cref{tab:rowhammer_defenses} provides an overview of Rowhammer defenses and the corresponding defense classes.
We defer a discussion of hardware-based defenses to \cref{sec:discussion_mitigations}.

In the following, we briefly describe the assumptions and implications for each of the defense classes, as well as an exhaustive list of defenses for each class.

\begin{table}[t]
  \setlength{\aboverulesep}{0pt}
  \setlength{\belowrulesep}{0pt}
  \caption{Rowhammer defenses for commodity systems.}\label{tab:rowhammer_defenses}
  \resizebox{0.96\hsize}{!}{
  \setlength\tabcolsep{1.5pt}
  \begin{tabular}{r llllllllllllllllll}
    \diagbox{Methodology}{Defense} &\,\, & \rotatebox{70}{MASCAT~\cite{Irazoqui2016mascat}} \hspace{-20em} & \rotatebox{70}{\citeA{Chiappetta2015}} \hspace{-20em} & \rotatebox{70}{\citeA{Zhang2016CloudRadar}} \hspace{-20em} & \rotatebox{70}{\citeA{Herath2015}} \hspace{-20em} & \rotatebox{70}{HexPADS~\cite{Payer2016}} \hspace{-20em} & \rotatebox{70}{\citeA{Gruss2016Flush}} \hspace{-20em} & \rotatebox{70}{ANVIL~\cite{Aweke2016}} \hspace{-20em} & \rotatebox{70}{\citeA{Corbet2016}} \hspace{-20em} & \rotatebox{70}{No OOM~\cite{VanDerVeen2016,Gruss2016Row}} \hspace{-20em} & \rotatebox{70}{G-CATT~\cite{Brasser2017catt}} \hspace{-20em} & \rotatebox{70}{B-CATT~\cite{Brasser2017catt}} \hspace{-20em} & \rotatebox{70}{TRR~\cite{LPDDR4}} \hspace{-20em} & \rotatebox{70}{MAC~\cite{LPDDR4}} \hspace{-20em} & \rotatebox{70}{PARA/CRA/PRA~\cite{Kim2014,Kim2015}} \hspace{-20em} & \rotatebox{70}{ARMOR \cite{Ghasempour2015}} \hspace{-20em} & \rotatebox{70}{ECC/Chipkill~\cite{Kim2014,Chipkill}} \hspace{-20em} & \rotatebox{70}{Refresh Rate~\cite{Kim2014}} \hspace{-20em} \\
    \toprule \vspace{-0.3cm}\\ 
 \multicolumn{18}{l}{\footnotesize{\textsc{Detection}}} \vspace{0.85em} \\
 \vspace{-2em} \\
    Static Analysis&		 		& \cmarkfull 	& \cmarkempty 	& \cmarkempty 	& \cmarkhalf 	& \cmarkempty 	& \cmarkempty 	& \cmarkempty 	& \cmarkempty 	& \cmarkempty 	& \cmarkempty 	& \cmarkempty 	& \cmarkempty 	& \cmarkempty 	& \cmarkempty 	& \cmarkempty 	& \cmarkempty  	& \cmarkempty 	\\
    Performance Counters& 			& \cmarkempty 	& \cmarkfull 	& \cmarkfull	& \cmarkfull	& \cmarkfull	& \cmarkfull	& \cmarkfull 	& \cmarkempty 	& \cmarkempty 	& \cmarkempty 	& \cmarkempty 	& \cmarkempty 	& \cmarkempty 	& \cmarkempty 	& \cmarkempty 	& \cmarkempty  	& \cmarkempty 	\\
    \vspace{0.4em} Memory Access Pattern& 			& \cmarkempty 	& \cmarkempty 	& \cmarkempty	& \cmarkempty	& \cmarkempty	& \cmarkempty	& \cmarkfull 	& \cmarkfull 	& \cmarkempty 	& \cmarkempty 	& \cmarkempty 	& \cmarkempty 	& \cmarkempty 	& \cmarkempty 	& \cmarkempty 	& \cmarkempty  	& \cmarkempty 	\\
\cdashline{1-19} \\ \vspace{-2em} \\
 \multicolumn{18}{l}{\footnotesize{\textsc{Neutralization}}} \vspace{0.85em} \\
 \vspace{-2em} \\
    Physical Proximity& 				& \cmarkempty 	& \cmarkempty 	& \cmarkempty 	& \cmarkempty 	& \cmarkempty 	& \cmarkempty 	& \cmarkempty 	& \cmarkempty 	& \cmarkempty 	& \cmarkfull 	& \cmarkempty 	& \cmarkempty 	& \cmarkempty 	& \cmarkempty 	& \cmarkempty 	& \cmarkempty  	& \cmarkempty 	\\
    \vspace{0.4em} Memory Footprint& 				& \cmarkempty 	& \cmarkempty 	& \cmarkempty 	& \cmarkempty 	& \cmarkempty 	& \cmarkempty 	& \cmarkempty 	& \cmarkempty 	& \cmarkfull 	& \cmarkempty 	& \cmarkempty 	& \cmarkempty 	& \cmarkempty 	& \cmarkempty 	& \cmarkempty 	& \cmarkempty  	& \cmarkempty 	\\
 \cdashline{1-19} \\ \vspace{-2em} \\
 \multicolumn{18}{l}{\footnotesize{\textsc{Elimination}}} \vspace{0.7em} \\
 \vspace{-2em} \\
    Bootloader& 				& \cmarkempty 	& \cmarkempty 	& \cmarkempty 	& \cmarkempty 	& \cmarkempty 	& \cmarkempty 	& \cmarkempty 	& \cmarkempty 	& \cmarkempty 	& \cmarkempty 	& \cmarkfull 	& \cmarkempty 	& \cmarkempty 	& \cmarkempty 	& \cmarkempty 	& \cmarkempty  	& \cmarkempty 	\\
    Hardware Modification& 			& \cmarkempty 	& \cmarkempty 	& \cmarkempty 	& \cmarkempty 	& \cmarkempty 	& \cmarkempty 	& \cmarkempty 	& \cmarkempty 	& \cmarkempty 	& \cmarkempty 	& \cmarkempty 	& \cmarkfull 	& \cmarkfull 	& \cmarkfull 	& \cmarkfull 	& \cmarkfull  	& \cmarkempty 	\\
    \vspace{0.4em} BIOS Update& 				& \cmarkempty 	& \cmarkempty 	& \cmarkempty 	& \cmarkempty 	& \cmarkempty 	& \cmarkempty 	& \cmarkempty 	& \cmarkempty 	& \cmarkempty 	& \cmarkempty 	& \cmarkempty 	& \cmarkempty 	& \cmarkempty 	& \cmarkempty 	& \cmarkempty 	& \cmarkempty 	& \cmarkfull 	\\
    \bottomrule
  \end{tabular}
  }
  \vspace{0.25em}
  \newline

   Symbols indicate whether a defense is part of defense class (\circletfill), optional aspects of the defense are part of a defense class (\circletfillhl), or a defense is not part of a defense class (\circlet).
\end{table}

\parhead{Static Analysis}\label{sec:cat:sa}
The underlying assumption of defenses based on static analysis (\textbf{D1}) is that the attack (binary) code can be accessed.
This defense class is especially interesting for offline analysis, \eg before adding software to an app store.
If the detection works, the user cannot be attacked anymore.
Static analysis is used by \citeA{Irazoqui2016mascat} in MASCAT, an automated static code analysis tool to detect microarchitectural attacks on a large scale.
\citeA{Herath2015} proposed to suspend programs with high cache miss rates and analyze instructions near the instruction pointer.
    
\parhead{Performance Counter Monitoring}\label{sec:cat:pca}
The underlying assumptions of defenses based on performance counter analysis (\textbf{D2}) are that the performance counters are available and that they include operations of the attacker program.
A typical parameter for Rowhammer detection is the number of cache hits and cache misses.
Detecting Rowhammer at runtime leaves a theoretical chance of missing an attack.
If the detection works, attacks are stopped before they can exploit a bit flip.
The use of performance counters is the basis of several defenses~\cite{Herath2015, Payer2016, Gruss2016Flush}.
The underlying \FlushReload loop of Rowhammer is also detected by cache attack defenses~\cite{Chiappetta2015,Zhang2016CloudRadar}.

\parhead{Memory Access Patterns Monitoring}\label{sec:cat:map}
The underlying assumptions of defenses based on memory access patterns (\textbf{D3}) are that Rowhammer attacks require a large number of cache misses on one row, and a large cumulative number of accesses on other rows in the same DRAM bank. Assuming this, Rowhammer attacks can be detected and stopped before they cause bit flips~\cite{Aweke2016,Corbet2016}.
ANVIL~\cite{Aweke2016} detects Rowhammer in two stages: First, it monitors the last-level cache miss ratio. Next, if the cache miss ratio exceeds a threshold, ANVIL uses Intel PEBS to monitor the addresses of cache misses and distinguish Rowhammer attacks from legitimate work loads.
For every candidate row, ``other row access samples from the same DRAM bank'' are checked (\cf Section 3.3 in~\cite{Aweke2016}). Only if there are enough accesses to other rows of the same bank, an attack is detected and victim rows are refreshed~\cite{Aweke2016}. 

\parhead{Preventing Physical Proximity}\label{sec:cat:ma}
The underlying assumption of defenses based on preventing physical proximity (\textbf{D4}) is that Rowhammer attacks need to flip bits in page tables or other kernel pages to take over the system.
A memory allocator can prevent physical proximity of user pages and kernel pages.
G-CATT~\cite{Brasser2017catt} is the only published defense in this class.
G-CATT isolates kernel pages from user pages by leaving a gap in physical memory.
If the isolation works, the user cannot take over the kernel and the system anymore.

\parhead{Memory Footprints}\label{sec:cat:mf}
The underlying assumptions of defenses based on prohibiting conspicuous memory footprints (\textbf{D5}) are that Rowhammer attacks need to allocate large amounts of memory to scan for bit flips and almost exhaust the entire memory to surgically place a page in a specific physical location to trigger and exploit a Rowhammer bit flip.
While the memory consumption of the attacker can already raise suspicion, both spraying~\cite{Seaborn2015BH,Gruss2016Row} and grooming~\cite{VanDerVeen2016} easily exhaust the entire memory in a way that gets the attacker process killed by the OS.
The memory allocator by default already avoids placing kernel pages near userspace pages, and it only deviates from this behavior in near-out-of-memory situations.
Not deviating from the default behavior to prevent adversarial memory exhaustion was mentioned in Rowhammer attack papers~\cite{VanDerVeen2016,Gruss2016Row}.
If the memory allocator prevents adversarial memory exhaustion, an attacker cannot force target pages to specific memory locations anymore.

\section{Attacker Model}\label{sec:threat_model}
Our attacker model makes the following fundamental assumptions about the hardware, the OS, installed defense mechanisms, and attacker capabilities:

\parhead{Hardware} The installed DRAM modules are susceptible to Rowhammer bit flips and no dedicated hardware-based Rowhammer defense mechanisms are in place.

\parhead{Operating System} The OS is up-to-date and fully patched, and no known software vulnerabilities exist that an attacker could exploit to elevate privileges.

\parhead{Defenses} The system is protected with state-of-the-art Rowhammer defenses.
Specifically, at least one defense from each defense class is deployed, including static analysis~\cite{Irazoqui2016mascat}, hardware performance counters~\cite{Herath2015, Payer2016, Gruss2016Flush, Aweke2016, Chiappetta2015,Zhang2016CloudRadar}, memory access pattern analysis~\cite{Aweke2016}, physical proximity prevention~\cite{Brasser2017catt}, and prevention of near-out-of-memory situations~\cite{VanDerVeen2016,Gruss2016Row}.

\parhead{Attacker Capabilities} We assume that an attacker can start an arbitrary unprivileged user program and that the attacker can launch an SGX enclave, which is also unprivileged.

\section{High-Level View of the Attacks}\label{sec:highlevel}

\begin{table}[t]
  \setlength{\aboverulesep}{0pt}
  \setlength{\belowrulesep}{0pt}
  \centering
  \caption{How the different defense classes are bypassed.}\label{tab:hlv_attack}
  \resizebox{\hsize}{!}{
  \begin{tabular}{rccccccc}
    \diagbox{Bypass}{Defense Class} &\,\, &
    \rotatebox{90}{\parbox{2.0cm}{Static Analysis}} &
    \rotatebox{90}{\parbox{2.0cm}{Performance Counters}} &
    \rotatebox{90}{\parbox{1.8cm}{Memory Access Pattern}} &
    \rotatebox{90}{\parbox{1.5cm}{Physical Proximity}} &
    \rotatebox{90}{\parbox{2.0cm}{Memory footprint}} & \\
    \toprule \vspace{-0.5em} \\
    Intel SGX                     & & \cmarkfull  & \cmarkfull  & \cmarkempty & \cmarkempty & \cmarkempty \\
    One-location hammering     & & \cmarkempty & \cmarkempty & \cmarkfull  & \cmarkempty & \cmarkempty \\
    Opcode flipping               & & \cmarkempty & \cmarkempty & \cmarkempty & \cmarkfull  & \cmarkempty \\
    \vspace{0.4em}
    Memory waylaying              & & \cmarkempty & \cmarkempty & \cmarkempty & \cmarkempty & \cmarkfull \\
    \midrule
	\textbf{Defense class defeated} & & \cmarkfull  & \cmarkfull  & \cmarkfull  & \cmarkfull  & \cmarkfull \\
    \bottomrule
  \end{tabular}
  }
\end{table}

In this section, we provide a high-level overview of the attack primitives we develop for our privilege-escalation attack in native environments and our denial-of-service attack in cloud environments, despite the presence of defenses from all defense classes from \cref{sec:threat_model}.
\cref{tab:hlv_attack} summarizes how we defeat every single defense class.

To defeat defense class \textbf{D1} (static analysis), we run our attack inside an SGX enclave.
Code in enclaves cannot be read or inspected, as the processor prevents all accesses to the enclave memory.
By encrypting the code and only decrypting it after the enclave is launched, a developer can hide arbitrary code within SGX enclaves.
Consequently, MASCAT~\cite{Irazoqui2016mascat} is incapable of detecting any microarchitectural or Rowhammer attack we perform inside the enclave.
Furthermore, the instruction stream cannot be searched for \texttt{clflush} instructions~\cite{Herath2015}.

Defense class \textbf{D2} (performance counters) is also defeated by running the attack inside an SGX enclave because the processor does not include SGX activity in process-specific performance counters for security reasons~\cite{Intel_SGX}.
Confirming this, Schwarz~\etal\cite{Schwarz2017SGX} observed that performance counters are not influenced by cache attacks running in SGX enclaves.
Hence, performance counters do not detect our attack.

\parhead{One-location Hammering}
To defeat defense class \textbf{D3} (memory access patterns), we introduce a new attack primitive, which we call \emph{one-location hammering}.
As older systems used an ``open-page'' memory controller policy where a memory row is kept open and buffered until the next memory row is accessed, double-sided and single-sided hammering cause frequent activations of rows by inducing cache misses on different rows of the same bank~\cite{Kim2014}.  Recently, however, modern systems employ more sophisticated memory controller policies, preemptively closing rows earlier than necessary, to optimize performance (\cf \cref{app:memory_controller_policies}).
We conjecture that this change in policy creates a  previously unknown Rowhammer effect, which we exploit with one-location hammering.

With one-location hammering, the attacker only runs a \FlushReload loop on a single memory address at the maximum frequency.
This continuously re-opens the same DRAM row, whenever the memory controller closes the row.
We observed that one-location hammering drains enough charge from the DRAM cells to induce bit flips.
As one-location hammering does not access different rows in the same bank, \textbf{D3} defenses, such as the second stage of ANVIL~\cite{Aweke2016}, do not detect the ongoing attack (\cf \cref{sec:cat:map}).
We describe one-location hammering in detail in \cref{sec:zero}.

\parhead{Opcode Flipping}
To defeat defense class \textbf{D4} (physical memory isolation), we introduce another new attack primitive, \emph{opcode flipping}.
All previous Rowhammer privilege-escalation attacks induced bit flips in carefully crafted page tables.
If the page table modification is successful, the attacker gains unrestricted read and write access to the physical memory, which is equivalent to having kernel privileges~\cite{Seaborn2015BH,Gruss2016Row,Xiao2016,VanDerVeen2016}.

With opcode flipping, we propose a novel way to exploit bit flips.
In the x86 instruction set, bit flips in opcodes yield other, in most cases, valid opcodes.
We show that with only a single targeted bit flip in an instruction, we can alter a (setuid) binary, \eg \texttt{sudo}, to provide an unprivileged process with root privileges.
As this is a bit flip in a user page, it breaks the underlying assumption of defense class \textbf{D4}, \ie G-CATT~\cite{Brasser2017catt}.

Previous attacks on unprivileged code~\cite{Seaborn2015BH} (\cf \cref{sec:bg_rh_att} for a detailed discussion) bypassed sandbox code sanitization by flipping bits in a bitmask used in a logical \texttt{and} in attacker-sprayed code.
In contrast to their work, we identify potential target bit flips in any opcode in a shared binary or library, modifying opcodes and the instruction stream.
Consequently, we illegitimately obtain root privileges by bypassing authentication checks. 
We detail opcode flipping in \cref{sec:opcode}.

\parhead{Memory Waylaying}
To defeat defense class \textbf{D5} (memory footprints), we introduce a novel alternative to memory spraying and grooming, called \emph{memory waylaying}.
Rowhammer attacks modify pages in a predictable way by placing them in physical memory locations where a known bit flips occur.
There are two techniques to achieve this: With \emph{spraying} the attacker fills the entire memory with copies of the generated data structure; with \emph{grooming} the attacker allocates the data structure to exploit in the exactly right moment.
Both methods require exhausting the entire memory and are easily detectable by monitoring memory consumption.
\emph{Memory waylaying} performs replacement-aware page cache eviction, using only page cache pages.
These pages are not visible in the system memory utilization as they can be evicted any time and hence, are considered as available memory.
Consequently, memory waylaying never causes the system to run out of memory.

We observed that page cache pages, after being discarded from DRAM, are loaded to a new random physical location upon access, on both Linux and Windows.
Through continuous eviction, the page is eventually placed on a vulnerable physical location.
Memory waylaying leverages the prefetch side-channel to detect when data in virtual memory is placed on a specific physical location.
By doing so, memory waylaying consumes a negligible amount of time and memory while waiting for the target page to be loaded to the target physical location.
Hence, it is difficult to detect.
Once the data is located at the desired position, the attacker hits it with the Rowhammer bit flip and exploits the modified binary to gain root privileges.
We describe memory waylaying in detail in \cref{sec:waylaying}.

\section{Opcode Flipping}\label{sec:opcode}
In this section, we describe \emph{opcode flipping}, a generic technique for exploiting bit flips in cached copies of binary files.
All previous generic Rowhammer privilege-escalation attacks (\ie obtaining root privileges) induced bit flips in the page number field of an attacker-generated page table, in order to change the memory page reference by some page table entry.
\citeA{Seaborn2015BH} (\cf \cref{sec:bg_rh_att} for a detailed discussion) bypassed sandbox code sanitization by flipping bits in a bitmask used in a logical \texttt{and} in attacker-sprayed code.

In contrast to previous work, we identify potential target bit flips in any opcode in a shared binary or library, modifying opcodes and the instruction stream.
In contrast to previous Rowhammer attacks based on memory spraying, the binary pages we attack cannot be sprayed and only exist a single time in the entire memory.
In order to find suitable bit flips  in system binaries, we used the following methodology.
First, we manually define ranges within in the binary for which bits could be flipped.
We then automatically test every single bit flip in these ranges, grouping the modified binaries by the result of their corresponding execution.
Finally we manually analyze the results, looking for devastating outcomes (such a obtaining root permissions without knowing the root password) and target these bits via our Rowhammer attack.

Opcode flipping exploits that bit flips in opcodes can yield other, yet valid, opcodes.
These opcodes are often very similar to the original opcode but have different, possibly inverted, semantics. 
One prerequisite of opcode flipping is the ability to flip a bit of a target binary page with surgical precision.
For now, we assume that the attacker can cause such a precise bit flip and discuss the effect of such bit flips, before we show in \cref{sec:waylaying} how a file can be placed in memory accordingly.

\parhead{Opcode Flipping Case Study}\label{sec:case-study}
To illustrate opcode flipping we consider the example of a single bit flip in the x86 opcode \texttt{JE = 0x74} (jump if equal).
A single bit flip in this opcode can yield the opcodes \texttt{JNE = 0x75} (jump if not equal), \texttt{JBE = 0x76} (jump if below or equal), \texttt{JO = 0x70} (jump if overflow), \texttt{JL = 0x7C} (jump if lower), \texttt{PUSHQ = 0x54} (push quad word), \texttt{XORB = 0x34} (xor byte), \texttt{HLT = 0xF4} (halt), and the prefix \texttt{0x64}.
Only 21 out of 255 two-byte sequences starting with the prefix \texttt{0x64} are illegal opcodes.

Similarly, flips in \texttt{TEST} instructions preceding a conditional jump have the same effect. 
For example, with a single bit flip, the instruction \texttt{TEST EAX,EAX}, which sets the zero flag if \texttt{EAX} is zero, can be transformed to \texttt{XCHG EAX,EAX}, which never modifies the zero flag. 
Tests and conditional jumps are used in virtually all computer programs, and they control the decision logic of the programs.
Therefore, we focus on flips in these instructions.
As we show, bit flips in such instructions are sufficient to achieve our goals. 

\parhead{Exploitable Opcodes in Real-World Binaries}\label{sec:exploitable-opcodes}
To exploit opcode flipping for privilege escalation, we target userspace applications with the \texttt{setuid} bit set, which are run as root. 
On Ubuntu 17.04, there are 16 \texttt{setuid} binaries owned by root, all being potential targets for privilege escalation using a bit flip. 
We analyzed one of the most prominent targets for privilege escalation, the \texttt{sudo} binary and \texttt{sudoers.so} shared library (henceforth \emph{\texttt{sudo} binary}).

We identified two regions in the \texttt{sudo} binary in which a bit flip can be exploited. 
First, the check whether the user is allowed to use \texttt{sudo}, \ie if the user is in the \texttt{sudoers} file. 
Second, the check whether the entered password is correct. 
In this work, we focus on the latter.

We located 29 different offsets in the binary where a bit flip breaks the password verification logic. 
All identified bit flips affect the test or the conditional jump of the password-verification location. 
Successful attacks on the conditional jump change the condition so that it treats an incorrect password as if it was correct.
Attacks on the test instruction result in different operations which ensure that the zero flag is clear, either by clearing it, \eg \texttt{ADD AL,0xC0}, or by maintaining the previous, clear, value.
We provide a list with offsets and their effect on the opcode at this position, in \cref{app:bitflips_sudo}.

As shown in the following section, bit flip positions in memory are uniformly distributed, allowing exploitation of any of the 29 offsets in the \texttt{sudo} binary to gain root privileges. 

\section{One-location Rowhammer}\label{sec:zero}
In this section, we describe the hammering technique we use to induce bit flips.
We assume that the attacker already knows exploitable bit offsets in binaries and only searches for memory locations where these bit offsets can be flipped through Rowhammer.
We propose one-location Rowhammer as a novel alternative technique based on previously unknown Rowhammer effects.
The scanning is performed from within the enclave and hence, cannot be observed through performance counters, source-code analysis or binary analysis.

Previous work described two different hammering techniques, double-sided hammering, and single-sided hammering, as described in more detail in \cref{sec:bg_rh_att}.

One-location hammering truly hammers only one memory location, \ie the attacker does not directly induce row conflicts but only re-opens one row permanently.
The core of one-location hammering is a \FlushReload loop hammering a single randomly chosen address, 
voiding the assumptions of defense class \textbf{D3}.
Both, one-location hammering and single-sided hammering are oblivious to virtual-to-physical address mappings.
Hence, we can also apply both hammering techniques if physical address mappings are not available.

\begin{figureA}[t]{bit_flip_comparison}[Flippable bit offsets over \SI{4}{\kilo B}-aligned memory regions for different hammering techniques. Bit flips from 0 to 1 (blue) and bit flips from 1 to 0 (red) may occur at any bit offset.]
\,\quad\,
\begin{subfigure}{0.25\hsize}
\centering
 \includegraphics[width=\hsize]{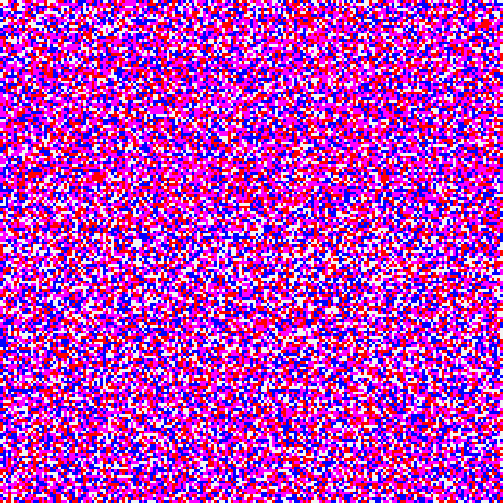}
\caption{Double-sided}
\end{subfigure}
\hfill
\begin{subfigure}{0.25\hsize}
\centering
 \includegraphics[width=\hsize]{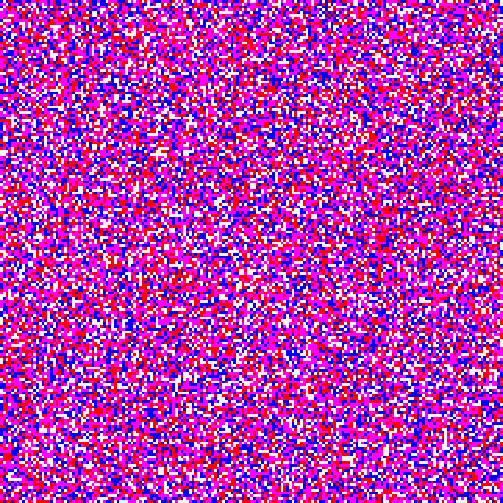}
\caption{Single-sided}
\end{subfigure}
\hfill
\begin{subfigure}{0.25\hsize}
\centering
 \includegraphics[width=\hsize]{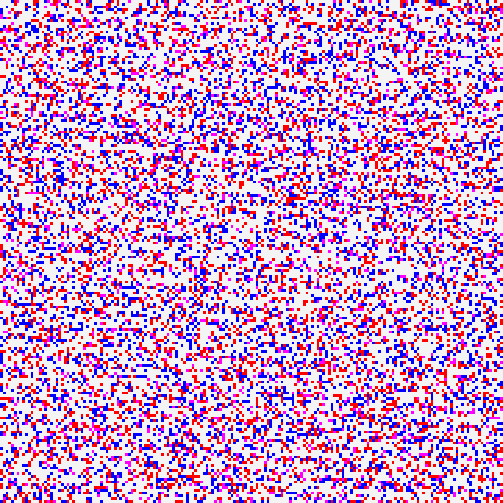}
\caption{One-location}
\end{subfigure}
\,\quad\,
\end{figureA}

We studied the distribution of bit flips over \SI{4}{\kilo B}-aligned memory regions, \ie pages, as this alignment can be obtained through our memory waylaying technique described in \cref{sec:waylaying}.
We performed our analysis on a Skylake i7-6700K with two \SI{8}{\giga B} Crucial DDR4-2133 DIMMs.
We tested each technique for eight hours and scanned for bit flips after each hammering attempt (\ie after \SIx{5000000} rounds of \FlushReload on two or one address, respectively).
Each hammering attempt hammers random memory locations (randomly-chosen offsets on more than \SIx{100000} randomly-chosen \SI{4}{\kilo B} pages).
\cref{fig:bit_flip_comparison} shows the distribution of bit flip offsets accumulated over \SI{4}{\kilo B}-aligned memory regions for double-sided hammering, single-sided hammering, and one-location hammering.
We observe that \SIx{25223} out of \SIx{32768} bit offsets (\SI{77.0}{\percent}) can be flipped using double-sided hammering on at least one \SI{4}{\kilo B}-offset.
\SI{51.7}{\percent} of the bit flips were from 0 to 1.

Single-sided hammering does not induce more bit flips than double-sided hammering.
However, regarding bit offsets, we observe an even slightly more uniform distribution for single-sided hammering, with \SIx{25722} bit offsets (\SI{78.5}{\percent}).
\SI{54.1}{\percent} of the bit flips were from 0 to 1.

One-location hammering only flipped \SIx{11969} out of \SIx{32768} bit offsets (\SI{36.5}{\percent}) on at least one \SI{4}{\kilo B}-offset.
\SI{51.6}{\percent} of the bit flips were from 0 to 1.
This is worse than double-sided hammering and single-sided hammering.
Still, our results show for the first time, that one-location hammering drains sufficient charge from the DRAM cells to induce bit flips.

We validated our results by reproducing them in a short series of tests on a Haswell i7-4790 with two Kingston DDR3-1600 DIMMs.
We observe bit flips for all hammering techniques, including one-location hammering.
On an Ivy Bridge i5-3230M with two Samsung DDR3-1600 SO-DIMMs we observe a significantly higher number of bit flips for double-sided hammering than for single-sided hammering, while bit flips from one-location hammering were rare and not reliably reproducible.
Our measurements indicate that this machine uses an open-page memory controller policy, as opposed to the more efficient policies used on the other two systems (\cf \cref{app:memory_controller_policies}).
However, bit flips from 0 to 1 and from 1 to 0 have approximately the same probability on all three systems.

Our data shows that the bit flips over pages generally follow a uniform distribution if a significant amount of memory is tested.
As our attacker aims at finding bit flips for specific offsets on \SI{4}{\kilo B} pages, the runtime of the bit flip templating phase depends on the number of exploitable bit flip offsets.
In case of the 29 bit offsets we found in \texttt{sudo}, the expected runtime on our Skylake system is less than 17 minutes per target bit flip for double-sided hammering, and less than 19 minutes for single-sided hammering.
With one-location hammering the expected runtime increases to 56 minutes until a target bit flip is found.
Hence, one-location hammering is $3.3$ times slower in finding the target bit flip than comparable hammering methods.
If evasion of defense class \textbf{D3} is a goal, a slow-down factor of $3.3$ is practical.

Deciding to run the stealthy templating longer than necessary, \ie searching for more than one bit flip, reduces the runtime of the waylaying phase (\cf \cref{sec:waylaying}) significantly, as the attacker learns more addresses suitable for the attack.

The templating only keeps the CPU core of the enclave busy but causes no other system utilization, \ie it does not exhaust memory, as we rely on the memory allocation of our waylaying technique, that we present in the following section.

\section{Memory Waylaying}\label{sec:waylaying}
The attacker knows which bit offsets in pages of binaries to target to obtain root privileges, and how to hammer physical memory locations to obtain a bit flip at the right bit offset.
The remaining problem is the inherent challenge of Rowhammer: Placing the target page at a physical location where the required bit flip can be induced.
The known approaches to solve this challenge are spraying, \ie filling the entire memory with copies of the page, or grooming, \ie allocating the target page in exactly the right moment~\cite{Yim2016}.
However, the page cache keeps every binary page only once in memory.
Linux prioritizes keeping binary pages in memory upon eviction.
Hence, spraying is not applicable in our attack and grooming would require out-of-memory situations to force eviction of the binary page.
In this section, we present \emph{memory waylaying}, a reliable approach to solving the challenge of memory placement.
It is a generic stealthy alternative to spraying and grooming, relying on a prediction oracle to determine whether a target page is at the right physical memory location.

In \cref{sec:prefetch-oracle}, we show how the prefetch side-channel attack~\cite{Gruss2016CCS} can be leveraged as an oracle. 
In \cref{sec:page-cache-eviction}, we present a technique to evict a target page from the page cache, forcing relocation at the next access. 
In \cref{sec:position-page}, we describe how the prefetch attack and the page cache eviction are combined to the stealthy memory waylaying.
We also present a fast variant, called \emph{memory chasing}, which sacrifices stealth for speed, with no sacrifice of reliability.

\subsection{Prefetch-based Prediction Oracle}\label{sec:prefetch-oracle}
In our memory waylaying attack, the attacker monitors page placement to detect mapping of one of the offsets in binaries and shared libraries to one of the target memory locations. 
We use the prefetch address-translation oracle~\cite{Gruss2016CCS} to perform this monitoring.
The oracle exploits the direct-physical mapping in the Linux kernelspace.
The prefetch address-translation oracle provides an attacker with the information whether two virtual addresses map to the same physical address, even in the presence of address-space layout randomization.

The address-translation oracle consists of two steps, a sequence of prefetch instructions and a \FlushReload attack, to measure the effect of the prefetch.
While the attack is prone to false negatives due to ignored prefetch instructions, the \FlushReload attack at its core has virtually no false positives~\cite{Yarom2014}, \ie there is no cache hit if the address was not actually cached.
While both steps can generally be executed in SGX enclaves, performing a \FlushReload attack requires highly accurate timing measurements.
On SGX2, \texttt{rdtsc} is available within enclaves.
On SGX1, Schwarz~\etal\cite{Schwarz2017SGX} demonstrated that accurate timing can be obtained by using counting threads and \citeA{Wang2017SGX} mirrored \texttt{rdtsc} into the enclave.
Our experiments with both approaches show that we can use either to obtain sufficiently accurate timing inside enclaves.

The address-translation oracle is first used in our attack to determine offsets in the direct-physical map with exploitable bit flips.
It is then used a second time, to continuously monitor the set of target addresses during the memory waylaying.
When an address match is detected, the next step of the attack is triggered, \ie hitting the target page with Rowhammer.

Our prefetch address-translation oracle, which we optimized for stability, experienced no false positives over a time frame of \SIx{3737} seconds and a true positive every \SIx{4.5} seconds, \ie the expected value for the true positive rate is \SI{50}{\percent} when measuring for \SIx{4.5} seconds. 
When optimized for performance we can achieve the same performance as \citeA{Gruss2016CCS}, \ie an expected measurement time of less than \SIx{50} milliseconds per address without false positives, but with a higher false negative rate. 
The search for the physical addresses is combined into one prefetch side-channel attack, \ie one prefetch operation and as many \FlushReload loops as page translations the attacker wants to find.
Hence, the runtime does not increase significantly with the number of addresses, but only linearly in the amount of system memory.

\subsection{Page Cache Eviction}\label{sec:page-cache-eviction}

Both on Windows and Linux, files are cached page-wise in the file page cache upon the first access to the corresponding page.
Any subsequent access to a page of a file is directly served from the page cache. 
Thus, one prerequisite for memory waylaying is a technique to deterministically evict a page of a file from the page cache. 
Eviction ensures that any subsequent access to the file cannot be served from the page cache anymore, and the file is mapped to a new physical location. 

Any unprivileged process could evict data from the page cache by simply allocating a large amount of memory, such that page cache pages must be evicted.
This is similar to the memory exhaustion techniques in previous Rowhammer attacks and risks system crashes due to out of memory situations~\cite{Seaborn2015BH,Gruss2016Row,VanDerVeen2016}.
We examined the behavior of the page cache replacement algorithm to find a more reliable way to trigger eviction.
While Linux provides privileged interfaces to do so, we need an approach which works without any privileges and from within enclaves, \ie only with regular memory accesses.

A fundamental observation we made is that the replacement algorithm of the Linux page cache prioritizes eviction of non-executable pages over executable pages.
However, it does evict executable pages when filling the page cache with \emph{read-only executable} pages.
On Windows, executable and non-executable file-backed pages can be used equally.
This forms a basic primitive that allows us to efficiently and reliably evict a selected page from the page cache.
Because the page cache only uses otherwise unused memory pages, the technique does not result in memory pressure and avoids the unresponsiveness and out-of-memory situations that memory exhaustion causes~\cite{Seaborn2015BH,Gruss2016Row,VanDerVeen2016}.

For both approaches, memory exhaustion and replacement-aware page cache eviction, the amount of data which has to be accessed is at most the total amount of main memory in the system. 
To evaluate how much memory has to be allocated for the eviction to be successful, we use the Linux \texttt{mincore} function.
The \texttt{mincore} function tells whether a given page is in the page cache.
An attacker could also use this function to optimize the page cache eviction during an attack, \ie abort the replacement-aware page cache eviction as soon as the page to be evicted is not in the page cache anymore. 
However, this is a trade-off between stealth and performance, as the OS can monitor calls to the \texttt{mincore} function.

\begin{figureA}[t]{eviction-exhaustion}[Our replacement-aware page cache eviction only leads to negligible memory increase, whereas existing techniques are close to an out-of-memory situation.]
\resizebox{\hsize}{!}{
    \begin{tikzpicture}[scale=0.85]
	\pgfplotsset{every axis legend/.append style={at={(1.25,1)},anchor=north}}
	\begin{axis}[
	legend columns=1,
	xlabel={Memory Usage [\%]},
	ylabel={\# Cases},
	height=3cm,
	width=\hsize,
	ybar,
	ymin=2,
	ymax=600,
	xmin=4500,
	xmax=13100,
	scaled x ticks=false,
	bar width=2pt,
	xmajorgrids=false,
	legend image code/.code={\draw[#1] (0.0cm,-0.2cm) rectangle (0.6cm,0.2cm);},
	xtick={4660,5948,...,12388},
	xticklabels={40,50,60,70,80,90,100}
	]
	\addplot+[blue,fill=blue!65,area legend] table[x=memory,y=evict] {data/eviction_mem.csv};
	\addlegendentry{Eviction};
	\addplot+[red,fill=red!65,area legend,postaction={pattern=north east lines}] table[x=memory,y=exhaust] {data/eviction_mem.csv};
	\addlegendentry{Exhaustion};
	\addplot+[green,fill=green,fill opacity=0.25,area legend,dashed] table[row sep=crcr] {
	  7385 0 \\
	  7389 600 \\
	  7394 0 \\
	};
	\addlegendentry{Before Attack};
	\addplot+[orange,fill=orange,fill opacity=0.25,area legend,draw=none,bar width=20pt] table[row sep=crcr] {
	  11278 0 \\
	  11279 600 \\
	  11270 0 \\
	};

	\pgfplotsset{
	    after end axis/.code={
		\node[orange,above,rotate=90] at (axis cs:12908,300){\small{OOM}};
	    }
	}

	\end{axis}
    \end{tikzpicture}
}
\end{figureA}

We evaluated our replacement-aware page cache eviction on an Intel Core i5-6200U with \SI{12}{\giga B} of main memory. 
For the experiment, we kept the system at an typical workload, namely a browser, a mail client, and a music player were running during the experiment. 
\cref{fig:eviction-exhaustion} compares traditional memory exhaustion with our replacement-aware page cache eviction to evict a specific page (in our experiment a page of the \texttt{sudo} binary) from the page cache. 
Our replacement-aware page cache eviction only incurs a slight increase of used memory, whereas the exhaustion-based technique is close to an out-of-memory situation. 
In \SI{0.78}{\percent} of our exhaustion tests, the test program was even terminated by the OS due to excessive memory usage. 
In contrast, our replacement-aware page cache eviction never leads to an out-of-memory situation. 
On average, for our replacement-aware page cache eviction, it was sufficient to access \SI{5544}{\mega B} of data to evict the target page of the \texttt{sudo} binary from the page cache. 
The replacement-aware page cache eviction takes on average \SIx{2.68} seconds. 
For higher workloads, an attacker has to access even less data to evict a specific page from the page cache, as the size of the page cache decreases with the memory usage of active applications. 
On Windows, the page cache eviction takes on average \SIx{10.10} seconds, as we cannot rely on the Linux \texttt{mincore} function to abort the eviction process. 

\subsection{Positioning Memory Pages}\label{sec:position-page}

\begin{figureA}[t]{waylaying-illustration}[Memory waylaying. In step (a) some pages are free (\protect\memFree). Our eviction (b) allocates all free pages for the page cache (\protect\memCache), but leaves occupied pages  (\protect\memSystem) untouched. Repeating the eviction, the target page $B$ (\protect\memBinary) is relocated, but the occupied memory remains the same. Eventually, $B$ is placed on the target physical location $X$ (\protect\memTarget) as illustrated (f).]
\adjustbox{max width=\hsize}{
 \begin{subfigureA}[t]{0.25\hsize}{wl_1}[Start]
 \centering
  \adjustbox{max width=0.99\hsize}{\begin{tikzpicture}[scale=0.5]
\usetikzlibrary{patterns}
\tikzstyle{system} = [postaction={draw},fill=blue]
\tikzstyle{cache} = [postaction={draw,pattern=north east lines, pattern color=red!50}, fill=black!10]
\tikzstyle{sudo} = [postaction={draw,pattern=north west lines, thick, pattern color=green!60!black}, fill=green!40]

\draw[system] (0,0) rectangle +(1,1);
\draw[cache] (1,0) rectangle +(1,1);
\draw[system] (2,0) rectangle +(1,1);
\draw[cache] (4,0) rectangle +(1,1);
\draw[cache] (5,0) rectangle +(1,1);
\draw[cache] (6,0) rectangle +(1,1);
\draw[sudo] (0,1) rectangle +(1,1) node[pos=.5] {\sffamily \textbf{B}};
\draw[cache] (1,1) rectangle +(1,1);
\draw[cache] (2,1) rectangle +(1,1);
\draw (3,1) rectangle +(1,1);
\draw (4,1) rectangle +(1,1);
\draw[system] (5,1) rectangle +(1,1);
\draw[system] (6,1) rectangle +(1,1);
\draw (0,2) rectangle +(1,1);
\draw (1,2) rectangle +(1,1);
\draw[cache] (2,2) rectangle +(1,1);
\draw[cache] (3,2) rectangle +(1,1);
\draw[cache] (4,2) rectangle +(1,1);
\draw (5,2) rectangle +(1,1);
\draw (6,2) rectangle +(1,1);
\draw (0,3) rectangle +(1,1);
\draw (1,3) rectangle +(1,1);
\draw[system] (2,3) rectangle +(1,1);
\draw[system] (3,3) rectangle +(1,1);
\draw[system] (4,3) rectangle +(1,1);
\draw[system] (5,3) rectangle +(1,1);
\draw[system] (6,3) rectangle +(1,1);
\draw[system] (0,4) rectangle +(1,1);
\draw[system] (1,4) rectangle +(1,1);
\draw[system] (2,4) rectangle +(1,1);
\draw (3,4) rectangle +(1,1);
\draw[cache] (4,4) rectangle +(1,1);
\draw (5,4) rectangle +(1,1);
\draw[cache] (6,4) rectangle +(1,1);
\draw[cache] (0,5) rectangle +(1,1);
\draw[cache] (1,5) rectangle +(1,1);
\draw (2,5) rectangle +(1,1);
\draw (3,5) rectangle +(1,1);
\draw[cache] (4,5) rectangle +(1,1);
\draw[cache] (5,5) rectangle +(1,1);
\draw (6,5) rectangle +(1,1);
\draw[system] (0,6) rectangle +(1,1);
\draw (1,6) rectangle +(1,1);
\draw (2,6) rectangle +(1,1);
\draw[system] (3,6) rectangle +(1,1);
\draw[system] (4,6) rectangle +(1,1);
\draw[cache] (5,6) rectangle +(1,1);
\draw[system] (6,6) rectangle +(1,1);
\draw[postaction={draw=red,ultra thick}] (3,0) rectangle +(1,1) node[pos=0.5] {\sffamily \textbf{X}};

\end{tikzpicture}}
  \vspace{-0.4em}
 \end{subfigureA}
 \begin{subfigureA}[t]{0.25\hsize}{wl_2}[Our Eviction]
 \centering
  \adjustbox{max width=0.99\hsize}{\begin{tikzpicture}[scale=0.5]
\usetikzlibrary{patterns}
\tikzstyle{system} = [postaction={draw},fill=blue]
\tikzstyle{cache} = [postaction={draw,pattern=north east lines, pattern color=red!50}, fill=black!10]
\tikzstyle{sudo} = [postaction={draw,pattern=north west lines, thick, pattern color=green!60!black}, fill=green!40]

\draw[system] (0,0) rectangle +(1,1);
\draw[cache] (1,0) rectangle +(1,1);
\draw[system] (2,0) rectangle +(1,1);
\draw[cache] (4,0) rectangle +(1,1);
\draw[cache] (5,0) rectangle +(1,1);
\draw[cache] (6,0) rectangle +(1,1);
\draw[cache] (0,1) rectangle +(1,1) ;
\draw[cache] (1,1) rectangle +(1,1);
\draw[cache] (2,1) rectangle +(1,1);
\draw[cache] (3,1) rectangle +(1,1);
\draw[cache] (4,1) rectangle +(1,1);
\draw[system] (5,1) rectangle +(1,1);
\draw[system] (6,1) rectangle +(1,1);
\draw[cache] (0,2) rectangle +(1,1);
\draw[cache] (1,2) rectangle +(1,1);
\draw[cache] (2,2) rectangle +(1,1);
\draw[cache] (3,2) rectangle +(1,1);
\draw[cache] (4,2) rectangle +(1,1);
\draw[cache] (5,2) rectangle +(1,1);
\draw[cache] (6,2) rectangle +(1,1);
\draw[cache] (0,3) rectangle +(1,1);
\draw[cache] (1,3) rectangle +(1,1);
\draw[system] (2,3) rectangle +(1,1);
\draw[system] (3,3) rectangle +(1,1);
\draw[system] (4,3) rectangle +(1,1);
\draw[system] (5,3) rectangle +(1,1);
\draw[system] (6,3) rectangle +(1,1);
\draw[system] (0,4) rectangle +(1,1);
\draw[system] (1,4) rectangle +(1,1);
\draw[system] (2,4) rectangle +(1,1);
\draw[cache] (3,4) rectangle +(1,1);
\draw[cache] (4,4) rectangle +(1,1);
\draw[cache] (5,4) rectangle +(1,1);
\draw[cache] (6,4) rectangle +(1,1);
\draw[cache] (0,5) rectangle +(1,1);
\draw[cache] (1,5) rectangle +(1,1);
\draw[cache] (2,5) rectangle +(1,1);
\draw[cache] (3,5) rectangle +(1,1);
\draw[cache] (4,5) rectangle +(1,1);
\draw[cache] (5,5) rectangle +(1,1);
\draw[cache] (6,5) rectangle +(1,1);
\draw[system] (0,6) rectangle +(1,1);
\draw[cache] (1,6) rectangle +(1,1);
\draw[cache] (2,6) rectangle +(1,1);
\draw[system] (3,6) rectangle +(1,1);
\draw[system] (4,6) rectangle +(1,1);
\draw[cache] (5,6) rectangle +(1,1);
\draw[system] (6,6) rectangle +(1,1);
\draw[postaction={draw=red,ultra thick}] (3,0) rectangle +(1,1) node[pos=0.5] {\sffamily \textbf{X}};

\end{tikzpicture}}
  \vspace{-0.4em}
 \end{subfigureA}
 \begin{subfigureA}[t]{0.25\hsize}{wl_3}[Access\\\phantom{(c)} Binary]
 \centering
  \adjustbox{max width=0.99\hsize}{\begin{tikzpicture}[scale=0.5]
\usetikzlibrary{patterns}
\tikzstyle{system} = [postaction={draw},fill=blue]
\tikzstyle{cache} = [postaction={draw,pattern=north east lines, pattern color=red!50}, fill=black!10]
\tikzstyle{sudo} = [postaction={draw,pattern=north west lines, thick, pattern color=green!60!black}, fill=green!40]

\draw[system] (0,0) rectangle +(1,1);
\draw[cache] (1,0) rectangle +(1,1);
\draw[system] (2,0) rectangle +(1,1);
\draw[cache] (4,0) rectangle +(1,1);
\draw[cache] (5,0) rectangle +(1,1);
\draw[cache] (6,0) rectangle +(1,1);
\draw[cache] (0,1) rectangle +(1,1) ;
\draw[cache] (1,1) rectangle +(1,1);
\draw[cache] (2,1) rectangle +(1,1);
\draw[cache] (3,1) rectangle +(1,1);
\draw[cache] (4,1) rectangle +(1,1);
\draw[system] (5,1) rectangle +(1,1);
\draw[system] (6,1) rectangle +(1,1);
\draw[cache] (0,2) rectangle +(1,1);
\draw[cache] (1,2) rectangle +(1,1);
\draw[cache] (2,2) rectangle +(1,1);
\draw[cache] (3,2) rectangle +(1,1);
\draw[cache] (4,2) rectangle +(1,1);
\draw[cache] (5,2) rectangle +(1,1);
\draw[cache] (6,2) rectangle +(1,1) ;
\draw[cache] (0,3) rectangle +(1,1);
\draw[cache] (1,3) rectangle +(1,1);
\draw[system] (2,3) rectangle +(1,1);
\draw[system] (3,3) rectangle +(1,1);
\draw[system] (4,3) rectangle +(1,1);
\draw[system] (5,3) rectangle +(1,1);
\draw[system] (6,3) rectangle +(1,1);
\draw[system] (0,4) rectangle +(1,1);
\draw[system] (1,4) rectangle +(1,1);
\draw[system] (2,4) rectangle +(1,1);
\draw[cache] (3,4) rectangle +(1,1);
\draw[cache] (4,4) rectangle +(1,1);
\draw[cache] (5,4) rectangle +(1,1);
\draw[sudo] (6,4) rectangle +(1,1) node[pos=.5] {\sffamily \textbf{B}};
\draw[cache] (0,5) rectangle +(1,1);
\draw[cache] (1,5) rectangle +(1,1);
\draw[cache] (2,5) rectangle +(1,1);
\draw[cache] (3,5) rectangle +(1,1);
\draw[cache] (4,5) rectangle +(1,1);
\draw[cache] (5,5) rectangle +(1,1);
\draw[cache] (6,5) rectangle +(1,1);
\draw[system] (0,6) rectangle +(1,1);
\draw[cache] (1,6) rectangle +(1,1);
\draw[cache] (2,6) rectangle +(1,1);
\draw[system] (3,6) rectangle +(1,1);
\draw[system] (4,6) rectangle +(1,1);
\draw[cache] (5,6) rectangle +(1,1);
\draw[system] (6,6) rectangle +(1,1);
\draw[postaction={draw=red,ultra thick}] (3,0) rectangle +(1,1) node[pos=0.5] {\sffamily \textbf{X}};

\end{tikzpicture}}
  \vspace{-0.4em}
 \end{subfigureA}
 \begin{subfigureA}[t]{0.25\hsize}{wl_4}[Repeat: Evict\\\phantom{(d)} + Access]
 \centering
  \adjustbox{max width=0.99\hsize}{\begin{tikzpicture}[scale=0.5]
\usetikzlibrary{patterns}
\tikzstyle{system} = [postaction={draw},fill=blue]
\tikzstyle{cache} = [postaction={draw,pattern=north east lines, pattern color=red!50}, fill=black!10]
\tikzstyle{sudo} = [postaction={draw,pattern=north west lines, thick, pattern color=green!60!black}, fill=green!40]

\draw[system] (0,0) rectangle +(1,1);
\draw[cache] (1,0) rectangle +(1,1);
\draw[system] (2,0) rectangle +(1,1);
\draw[cache] (4,0) rectangle +(1,1);
\draw[cache] (5,0) rectangle +(1,1);
\draw[cache] (6,0) rectangle +(1,1);
\draw[cache] (0,1) rectangle +(1,1) ;
\draw[cache] (1,1) rectangle +(1,1);
\draw[cache] (2,1) rectangle +(1,1);
\draw[cache] (3,1) rectangle +(1,1);
\draw[cache] (4,1) rectangle +(1,1);
\draw[system] (5,1) rectangle +(1,1);
\draw[system] (6,1) rectangle +(1,1);
\draw[cache] (0,2) rectangle +(1,1);
\draw[cache] (1,2) rectangle +(1,1);
\draw[cache] (2,2) rectangle +(1,1);
\draw[cache] (3,2) rectangle +(1,1);
\draw[cache] (4,2) rectangle +(1,1);
\draw[cache] (5,2) rectangle +(1,1);
\draw[cache] (6,2) rectangle +(1,1) ;
\draw[cache] (0,3) rectangle +(1,1);
\draw[cache] (1,3) rectangle +(1,1);
\draw[system] (2,3) rectangle +(1,1);
\draw[system] (3,3) rectangle +(1,1);
\draw[system] (4,3) rectangle +(1,1);
\draw[system] (5,3) rectangle +(1,1);
\draw[system] (6,3) rectangle +(1,1);
\draw[system] (0,4) rectangle +(1,1);
\draw[system] (1,4) rectangle +(1,1);
\draw[system] (2,4) rectangle +(1,1);
\draw[cache] (3,4) rectangle +(1,1);
\draw[cache] (4,4) rectangle +(1,1);
\draw[cache] (5,4) rectangle +(1,1);
\draw[cache] (6,4) rectangle +(1,1);
\draw[cache] (0,5) rectangle +(1,1);
\draw[cache] (1,5) rectangle +(1,1);
\draw[cache] (2,5) rectangle +(1,1);
\draw[cache] (3,5) rectangle +(1,1);
\draw[cache] (4,5) rectangle +(1,1);
\draw[cache] (5,5) rectangle +(1,1);
\draw[cache] (6,5) rectangle +(1,1);
\draw[system] (0,6) rectangle +(1,1);
\draw[cache] (1,6) rectangle +(1,1);
\draw[cache] (2,6) rectangle +(1,1);
\draw[system] (3,6) rectangle +(1,1);
\draw[system] (4,6) rectangle +(1,1);
\draw[sudo] (5,6) rectangle +(1,1) node[pos=.5] {\sffamily \textbf{B}};
\draw[system] (6,6) rectangle +(1,1);
\draw[postaction={draw=red,ultra thick}] (3,0) rectangle +(1,1) node[pos=0.5] {\sffamily \textbf{X}};

\end{tikzpicture}}
  \vspace{-0.4em}
 \end{subfigureA}
 \begin{subfigureA}[t]{0.25\hsize}{wl_5}[Repeat: Evict\\\phantom{(e)}  + Access]
 \centering
  \adjustbox{max width=0.99\hsize}{\begin{tikzpicture}[scale=0.5]
\usetikzlibrary{patterns}
\tikzstyle{system} = [postaction={draw},fill=blue]
\tikzstyle{cache} = [postaction={draw,pattern=north east lines, pattern color=red!50}, fill=black!10]
\tikzstyle{sudo} = [postaction={draw,pattern=north west lines, thick, pattern color=green!60!black}, fill=green!40]

\draw[system] (0,0) rectangle +(1,1);
\draw[cache] (1,0) rectangle +(1,1);
\draw[system] (2,0) rectangle +(1,1);
\draw[cache] (4,0) rectangle +(1,1);
\draw[cache] (5,0) rectangle +(1,1);
\draw[cache] (6,0) rectangle +(1,1);
\draw[cache] (0,1) rectangle +(1,1) ;
\draw[cache] (1,1) rectangle +(1,1);
\draw[cache] (2,1) rectangle +(1,1);
\draw[cache] (3,1) rectangle +(1,1);
\draw[cache] (4,1) rectangle +(1,1);
\draw[system] (5,1) rectangle +(1,1);
\draw[system] (6,1) rectangle +(1,1);
\draw[cache] (0,2) rectangle +(1,1);
\draw[cache] (1,2) rectangle +(1,1);
\draw[cache] (2,2) rectangle +(1,1);
\draw[cache] (3,2) rectangle +(1,1);
\draw[cache] (4,2) rectangle +(1,1);
\draw[cache] (5,2) rectangle +(1,1);
\draw[cache] (6,2) rectangle +(1,1) ;
\draw[cache] (0,3) rectangle +(1,1);
\draw[cache] (1,3) rectangle +(1,1);
\draw[system] (2,3) rectangle +(1,1);
\draw[system] (3,3) rectangle +(1,1);
\draw[system] (4,3) rectangle +(1,1);
\draw[system] (5,3) rectangle +(1,1);
\draw[system] (6,3) rectangle +(1,1);
\draw[system] (0,4) rectangle +(1,1);
\draw[system] (1,4) rectangle +(1,1);
\draw[system] (2,4) rectangle +(1,1);
\draw[cache] (3,4) rectangle +(1,1);
\draw[cache] (4,4) rectangle +(1,1);
\draw[cache] (5,4) rectangle +(1,1);
\draw[cache] (6,4) rectangle +(1,1);
\draw[cache] (0,5) rectangle +(1,1);
\draw[cache] (1,5) rectangle +(1,1);
\draw[sudo] (2,5) rectangle +(1,1)  node[pos=.5] {\sffamily \textbf{B}};
\draw[cache] (3,5) rectangle +(1,1);
\draw[cache] (4,5) rectangle +(1,1);
\draw[cache] (5,5) rectangle +(1,1);
\draw[cache] (6,5) rectangle +(1,1);
\draw[system] (0,6) rectangle +(1,1);
\draw[cache] (1,6) rectangle +(1,1);
\draw[cache] (2,6) rectangle +(1,1);
\draw[system] (3,6) rectangle +(1,1);
\draw[system] (4,6) rectangle +(1,1);
\draw[cache] (5,6) rectangle +(1,1);
\draw[system] (6,6) rectangle +(1,1);
\draw[postaction={draw=red,ultra thick}] (3,0) rectangle +(1,1) node[pos=0.5] {\sffamily \textbf{X}};

\end{tikzpicture}}
  \vspace{-0.4em}
 \end{subfigureA}
 \begin{subfigureA}[t]{0.25\hsize}{wl_6}[Stop if target\\\phantom{(f)} reached]
 \centering
  \adjustbox{max width=0.99\hsize}{\begin{tikzpicture}[scale=0.5]
\usetikzlibrary{patterns}
\tikzstyle{system} = [postaction={draw},fill=blue]
\tikzstyle{cache} = [postaction={draw,pattern=north east lines, pattern color=red!50}, fill=black!10]
\tikzstyle{sudo} = [postaction={draw,pattern=north west lines, thick, pattern color=green!60!black}, fill=green!40]

\draw[system] (0,0) rectangle +(1,1);
\draw[cache] (1,0) rectangle +(1,1);
\draw[system] (2,0) rectangle +(1,1);
\draw[cache] (4,0) rectangle +(1,1);
\draw[cache] (5,0) rectangle +(1,1);
\draw[cache] (6,0) rectangle +(1,1);
\draw[cache] (0,1) rectangle +(1,1) ;
\draw[cache] (1,1) rectangle +(1,1);
\draw[cache] (2,1) rectangle +(1,1);
\draw[cache] (3,1) rectangle +(1,1);
\draw[cache] (4,1) rectangle +(1,1);
\draw[system] (5,1) rectangle +(1,1);
\draw[system] (6,1) rectangle +(1,1);
\draw[cache] (0,2) rectangle +(1,1);
\draw[cache] (1,2) rectangle +(1,1);
\draw[cache] (2,2) rectangle +(1,1);
\draw[cache] (3,2) rectangle +(1,1);
\draw[cache] (4,2) rectangle +(1,1);
\draw[cache] (5,2) rectangle +(1,1);
\draw[cache] (6,2) rectangle +(1,1) ;
\draw[cache] (0,3) rectangle +(1,1);
\draw[cache] (1,3) rectangle +(1,1);
\draw[system] (2,3) rectangle +(1,1);
\draw[system] (3,3) rectangle +(1,1);
\draw[system] (4,3) rectangle +(1,1);
\draw[system] (5,3) rectangle +(1,1);
\draw[system] (6,3) rectangle +(1,1);
\draw[system] (0,4) rectangle +(1,1);
\draw[system] (1,4) rectangle +(1,1);
\draw[system] (2,4) rectangle +(1,1);
\draw[cache] (3,4) rectangle +(1,1);
\draw[cache] (4,4) rectangle +(1,1);
\draw[cache] (5,4) rectangle +(1,1);
\draw[cache] (6,4) rectangle +(1,1);
\draw[cache] (0,5) rectangle +(1,1);
\draw[cache] (1,5) rectangle +(1,1);
\draw[cache] (2,5) rectangle +(1,1);
\draw[cache] (3,5) rectangle +(1,1);
\draw[cache] (4,5) rectangle +(1,1);
\draw[cache] (5,5) rectangle +(1,1);
\draw[cache] (6,5) rectangle +(1,1);
\draw[system] (0,6) rectangle +(1,1);
\draw[cache] (1,6) rectangle +(1,1);
\draw[cache] (2,6) rectangle +(1,1);
\draw[system] (3,6) rectangle +(1,1);
\draw[system] (4,6) rectangle +(1,1);
\draw[cache] (5,6) rectangle +(1,1);
\draw[system] (6,6) rectangle +(1,1);
\draw[sudo,postaction={draw=red,ultra thick}] (3,0) rectangle +(1,1) node[pos=.5] {\sffamily \textbf{B}} node[pos=0.5] {\sffamily \textbf{X}} ;

\end{tikzpicture}}
  \vspace{-0.4em}
 \end{subfigureA}
}
\end{figureA}

We combine the prefetch translation oracle (\cf \cref{sec:prefetch-oracle}) and the replacement-aware page cache eviction (\cf \cref{sec:page-cache-eviction}) to maneuver a target page on one of the physical locations with a bit flip (\cf \cref{sec:zero}).
As an extension to memory waylaying, which is slow but stealthy, we also propose \emph{memory chasing}, a faster non-stealthy variant.

Both memory waylaying and memory chasing, leverage the prefetch translation oracle to test whether our exploitable page is at the correct (\ie vulnerable) physical page. 
As the physical page usually does not change often (\ie only if there is high memory pressure or the system is rebooted), memory waylaying periodically evicts the page cache. 
On a subsequent access to the target page, the access cannot be served from the page cache anymore, and a new physical page is allocated and mapped. 
This procedure works the same way on Windows and Linux, as illustrated in \cref{fig:waylaying-illustration}.

\begin{figureA}[t]{waylaying-heatmap}[Distribution of placements of a page in the physical memory of our test system (\SI{12}{\giga B}). Each square represents \SI{4}{\mega B}. Hatched (red) areas are unavailable to the system (\eg graphics memory). The darker (blue) an area, the more physical pages were in this area. Even a small number of relocations covers most of the physical memory.]
\vspace{-1mm}
 \includegraphics[width=0.9\hsize]{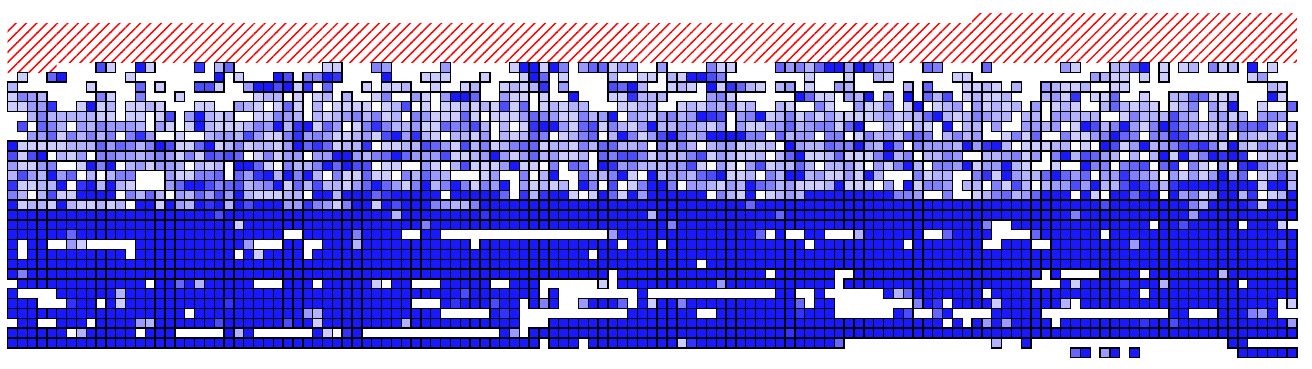}
\end{figureA}

We evaluate the distribution of physical page numbers used for a specific binary page on one of our test systems, an Intel Core i5 with \SI{12}{\giga B} of main memory. 
We repeated the memory waylaying process \SIx{57000} times, \ie the binary page was relocated \SIx{57000} times. 
Out of these \SIx{57000} relocations, we found \SIx{46720} unique physical page numbers, \ie the probability of maneuvering the binary to a physical location where it was already is only \SI{18}{\percent} after \SIx{57000} tries.
\cref{fig:waylaying-heatmap} visualizes the distribution of the \SIx{57000} relocations in physical memory.
We observe that even the small number of relocations we tested (\ie \SI{1.8}{\percent} of all pages) covered most of the physical memory, with the exception of occupied memory regions.
Thus, eventually the target binary page is placed at a physical memory location where the intended bit flip can be induced. 

The advantage of memory waylaying over conventional techniques, such as grooming or spraying, is that it is stealthy, as it does not exhaust the memory. 
The OS page cache is designed to occupy any unused page in the system.
Most pages are rarely accessed, but it is still more efficient to keep them in memory than to reload them from the disk.
Memory waylaying exploits this design, and as a consequence, it has no impact on memory utilization and only negligible impact on the overall system performance, as the page cache simply keeps a different set of pages in the otherwise unused memory.
In \cref{sec:exploit}, we detail the runtime of the waylaying phase in a practical example.

The disadvantage of memory waylaying is that the runtime can vary widely, from a few hours up to a few days, until the target page is placed on the correct physical location.
As a faster solution, we propose memory chasing, an adaption of memory waylaying which sacrifices stealth for speed.
Instead of waiting for the target page to be placed on a different physical page, we actively ``chase'' the binary in physical memory until it is at the correct physical page. 
Memory chasing runs outside of the enclave as it has a stronger interaction with userspace library functions.
To change the physical page of a target binary, memory chasing exploits the copy-on-write effect of \texttt{fork} as follows: 
\begin{compactenum}
 \item \texttt{mmap} the binary as private and \emph{writable}. 
 \item Fork the current process. 
 \item In the child process, write to the mapped binary. 
       This ensures that the page is copied to a new physical page. 
 \item Kill the parent process to release the old physical page. 
 \item Repeat until the page is at the intended physical location (check using the prefetch translation oracle)
\end{compactenum}

Although the binary content is now at the correct physical location, the page cache still holds the first version of the binary page, as the current page is \emph{dirty} (\ie modified). 
Thus, we have to trick the kernel into replacing the old binary page with the current one. 
We do this by evicting the page cache as described in \cref{sec:page-cache-eviction}. 
This removes the old (cached) binary page from the page cache. 
After the page cache is evicted, we unmap the current binary page and immediately map it again, however, this time with read-only and execute permissions. 
This ensures that the freed physical page is used to cache the binary in the page cache. 

Memory chasing is considerably faster than memory waylaying, as the page cache has to be evicted only once. 
Moving the physical page with memory chasing takes on average only \SI{36.7}{\micro\second}, whereas memory waylaying requires \SI{2.68}{\second}. 
On Windows, we could not test memory chasing as there is no equivalent to the \texttt{fork} function. 
With \SI{10.10}{\second}, memory waylaying requires slightly more time on Windows. 
However, both techniques have the advantage of not exhausting the memory in contrast to memory spraying and grooming. 
One disadvantage of memory chasing is the large number of fork system calls, occupying one CPU core.
Therefore, depending on how stealthy the attack must be, the attacker chooses which of the two primitives to use for reliable page cache eviction.
In \cref{sec:exploit}, we detail the runtime of memory chasing in a practical example.

\section{Evaluation of Attacks in Native and Cloud Environments}\label{sec:attacks}
In this section, we summarize our attacks and evaluate them in practical scenarios.
We first consider a cloud scenario with a simple attack, where an attacker is able to run our attack in virtual machines on multiple cloud servers.
We then consider a local scenario with our full attack, where an attacker is able to run our attack on personal computers and performs a privilege-escalation attack.
We detail the procedural steps of the attacks as well as the corresponding runtime.

\subsection{Abusing SGX for Denial-of-Service Attacks in the Cloud}\label{sec:dos}
Cloud servers are typically less susceptible to Rowhammer bit flips due to the presence of ECC, double refresh rates, and slower DRAM modules~\cite{Pessl2016}.
In the cloud scenario, the attacker uses our attack to identify vulnerable servers and take these servers down in a coordinated and distributed attack, \ie a denial-of-service attack.
In this attack, we do not aim for privilege escalation and hence, neither perform opcode flipping nor memory waylaying.
The attacker runs an unprivileged SGX enclave to evade defense classes \textbf{D1} and \textbf{D2}.

If, as discussed in \cref{sec:sgx}, an attacker induces bit flips in the encrypted memory area (EPC) of SGX, the CPU locks the memory controller (potentially incurring data corruption), causing the system to halt until it is rebooted manually.
Note that only a tiny fraction of \SI{4}{\kilo B} pages are adjacent to the \SI{128}{\mega B} EPC memory area.
For instance, on a system with \SI{16}{\giga B} dual-channel dual-rank DDR4 memory, only 256 pages (\SI{0.006}{\percent} of all pages) are in an adjacent DRAM row.
As different allocation mechanisms are used to allocate EPC pages and normal world pages, the attacker cannot accidentally hammer EPC addresses.
Hence, it is extremely unlikely to accidentally flip a bit in the EPC memory region.

Many cloud providers use KVM~\cite{Habib2008} or Xen~\cite{Barham2003} to run multiple virtual machines of different tenants in parallel on the same physical hardware.
To expose SGX features to virtual machines, Intel published the necessary kernel patches~\cite{KVMSGX, QemuSGX, XenSGX}.
Recently, \citeA{AzureSGX} introduced \emph{Azure confidential computing} that enables developers to use SGX in their cloud.

Our ``distributed'' denial-of-service attack consists of two phases, seek and destroy:
\begin{itemize}[nolistsep,align=left, leftmargin=9pt, labelwidth=0pt, itemindent=!]
\item \parhead{Seek} The attacker launches the attack enclave on many hosts in the cloud (\ie ``distributed''), and templates the DRAM for possible bit flips.
The runtime of this phase is in the range of multiple hours.
As the position of bit flips is uniformly distributed, an attacker learns from any bit flip while templating, that the DRAM very likely also vulnerable to bit flips in the EPC region used by SGX.
\item \parhead{Destroy} The attacker shuts down every vulnerable machine found in phase 1, by simultaneously triggering bit flips in EPC memory.
The runtime of this phase is in the range of seconds to minutes.
\end{itemize}

Besides ethical considerations on performing this experiment on a public cloud provider, we also found that no public cloud provider offers SGX support.
Microsoft's \emph{Azure confidential computing}~\cite{AzureSGX} can only be used as an early access program, that we have not been granted access to.
Instead, we performed the first part of our experiment on a dual CPU server system with two Intel Haswell-EP Xeon E5-2630 v3, a setup commonly found in public clouds.
We equipped the system with two Crucial DDR4-2133 DIMMs known to be susceptible to Rowhammer bit flips.
Our experiments showed that due to the significantly lower clock frequency (\SIx{60}--\SI{76}{\percent} of the clock frequency of an Intel Skylake i7-6700K) and the by-default doubled refresh rate, bit flips are much rarer.
Specifically, we observed only 3 bit flips in an 8 hour test.
However, this is sufficient for our denial-of-service attack.

In the second phase, our Rowhammer enclave starts to simultaneously hammer DRAM rows in the EPC on all hosts.
By triggering a bit flip within this memory region, the machine locks the memory controller (potentially incurring data corruption) and causes the system to halt until reboot.

As our Intel Haswell-EP system does not support Intel SGX, we performed the second part of our practical analysis on an Intel Skylake i7-6700K.
We verified that we are able to reproducibly crash the system within 10 seconds when hammering DRAM rows used by the EPC, as Intel SGX locked down the memory controller, halting the system and forcing us to power off the system manually.
We observed that occasionally, after powering on the system again, the system did not boot beyond the BIOS for several minutes.
After powering the system off and on again another time, the system regularly booted again.

Our results show that SGX introduces a significant security risk for cloud providers, allowing an attacker to cause hard-to-trace denial-of-service attacks and coordinated simultaneous take-down of multiple cloud servers, \eg in the \emph{Azure confidential computing} cloud~\cite{AzureSGX}.
As the attack hurts the availability and reliability of the cloud provider, it is especially interesting for parties with conflicting economic interests.

While the same attack could also be applied to a large number of personal computers, it is unclear how an attacker would profit from denial-of-service attacks on personal computers, especially in the face of the full privilege-escalation attack we detail in the next subsection.

In a concurrent independent work, Jang~\etal\cite{Jang2017SGXBomb} propose a similar attack, making the same observations as we did:
the system reset does not work properly following bit flips in SGX; any bit flip in the \SI{128}{\mega B} region causes the system to halt, making the attack easier than other Rowhammer attacks; all detection mechanisms are bypassed by hiding the Rowhammer code inside an enclave; and that just locking down the processor in case of a bit flip might not be the best defense scheme.
As a defense, they propose that future work should investigate whether there are non-process-specific performance counters which allow detection of suspicious activity in SGX enclaves.

\subsection{Abusing SGX to Hide Privilege-Escalation Attacks}\label{sec:exploit}
Personal computers are more susceptible to Rowhammer bit flips, as they usually are not equipped with ECC-RAM.
In this scenario, the attacker uses our full attack for privilege escalation from a regular unprivileged process to root privileges.
The crucial building blocks of this attack are opcode flipping and memory waylaying.
The attacker runs an unprivileged SGX enclave to evade defense classes \textbf{D1} and \textbf{D2}.

In our example attack, we apply opcode flipping (\cf \cref{sec:opcode}) to exploit bit flips in opcodes in the \texttt{sudo} binary of an up-to-date Ubuntu distribution.
Bit flips at some offsets in the binary (\cref{sec:exploitable-opcodes}) cause a skipping of authentication checks and, thus, provide us with root privileges.

The local attack requires two preparation steps:
\begin{itemize}[nolistsep,align=left, leftmargin=9pt, labelwidth=0pt, itemindent=!]
\item \parhead{Offline Preparation} The attacker determines which bit flip offsets in standard system executable binaries and shared libraries are exploitable.
This step is repeated for a large number of binaries and shared libraries of different distributions and versions.
The result of the offline preparation is a database of files, versions, and bit flip offsets (\cf \cref{sec:opcode}).
In this phase, we identified 29 exploitable bit offsets in \texttt{sudo}.
\item \parhead{Online Preparation} The attacker verifies that the binary and library versions on the target systems are in the database.
This is very likely the case if the victim uses a default installation of a popular Linux distribution, \eg Ubuntu, as all binaries and libraries are pre-compiled and hence, identical on virtually every installation.
\end{itemize}

\noindent After the preparation steps are completed, the attacker continues with the main attack, consisting of four phases:
\begin{itemize}[nolistsep,align=left, leftmargin=9pt, labelwidth=0pt, itemindent=!]
\item \parhead{Templating phase}
Our Rowhammer enclave templates memory for bit flips.
This is done via single-sided hammering or one-location hammering (\cf \cref{sec:zero}), which both are oblivious to physical addresses and hence, perfectly suited to be run in our Rowhammer enclave.
To defeat defense class \textbf{D3}, the attacker can use one-location hammering.
The memory is allocated via memory-mapped files (\cf \cref{sec:waylaying}), causing no significant increase in the resident memory and, thus, avoiding out-of-memory situations.

\begin{table}[t]
  \setlength{\aboverulesep}{0pt}
  \setlength{\belowrulesep}{0pt}
  \centering
  \caption{Optimal parameters and runtime of the attack.}\label{tab:runtime}
  \resizebox{\hsize}{!}{
  \begin{tabular}{rrrrr}
    \toprule
    \thead{Method} & \thead{Bitflips} & \thead{Templating} & \thead{Waylaying} & \thead{Total} \\
    \midrule
    Double-sided, waylaying & 91 & \SI{26.1}{h} &  \SI{69.4}{h} & \SI{95.5}{h} \\
    Single-sided, waylaying & 87 & \SI{27.5}{h} &  \SI{70.6}{h} & \SI{98.1}{h} \\
    One-location, waylaying & 50 & \SI{47.3}{h} &  \SI{90.5}{h} & \SI{137.8}{h} \\
\cdashline{1-5} \\ \vspace{-2em} \\
    Double-sided, chasing   & 1 & \SI{0.7}{h} &  \SI{43.7}{h} & \SI{44.4}{h} \\
    Single-sided, chasing   & 1 & \SI{0.7}{h} &  \SI{43.7}{h} & \SI{44.4}{h} \\
    One-location, chasing   & 1 & \SI{1.3}{h} &  \SI{44.0}{h} & \SI{45.4}{h} \\
    \bottomrule
  \end{tabular}
  }
\end{table}

The runtime of the templating phase and the waylaying phase pose an optimization problem (see \cref{app:attack_runtime}).
\cref{tab:runtime} shows the optimal solution for our scenario, \eg the runtime with one-location hammering is $47.3$ hours if followed by waylaying, and $1.3$ hours if followed by memory chasing.
Interruptions during this time frame are no problem, as the attacker tests independent memory locations and does not lose data over interruptions.
During the templating, the enclave occupies one CPU core, which is visible to the OS but which could also be explained by completely benign enclave operations.
The result of the templating phase is a list of physical pages with bit flips matching those from the preparation phase.
\item \parhead{Waylaying phase}
Our Rowhammer enclave uses a side channel to wait until one of the vulnerable target binary or library pages is placed on one of the exploitable memory locations (\cf \cref{sec:waylaying}). 
The prefetch-based prediction oracle tells us when the page has been loaded at the correct position.
Next, then we flip the bit in the opcode using one-location hammering in the \emph{hammering phase}.

The runtime of the waylaying phase depends on the number of bit flips found in the templating phase.
\cref{tab:runtime} shows the optimal solution for our scenario, \eg the runtime with one-location hammering is $90.5$ hours for memory waylaying and $44.0$ hours for memory chasing.
The result of the waylaying phase is that a target binary page is placed on the right physical page to trigger a predictable bit flip.
\item \parhead{Hammering phase}
The hammering phase only takes a few milliseconds, as it only induces the predictable bit flip on the target page using Rowhammer.
The attacker can verify whether a bit was flipped by reading the content of the binary page.
Thus, the result of the hammering phase is an unauthorized modification of the target binary, \ie in our case a malicious \texttt{sudo} binary.
\item \parhead{Exploitation phase}
As the binary page in memory now contains the modified opcodes, the privilege check in the target binary, \ie \texttt{sudo}, is circumvented.
Hence, the attacker simply runs the attacked binary and, thus, obtains root privileges.
Consequently, the exploitation phase also has a negligible runtime.
\end{itemize}

\noindent We performed all attack steps on an i7-6700K, showing that the attack can be mounted in practice.
Furthermore, we validated the templating on two other systems, an i5-3230M with Samsung DDR3-1600 memory, and an i7-4790 with Kingston DDR3-1600 memory.
We also validated the waylaying phase by running it for several days as a background process on a second machine (an i5-6200U), confirming that the user does not notice any attack activity and that it does not cause any system crashes.
To eliminate traces or avoid potential instabilities due to the binary modifications, an attacker can restore the unmodified binary page by simply evicting the page cache once more.
Upon the next access, the unmodified version is reloaded from the disk.

Our attack shows that existing countermeasures for commodity systems are incomplete and fundamental assumptions need to be refined to design effective countermeasures.

\section{Discussion}\label{sec:discussion}
In this section, we discuss limitations of our approach and additional observations we made while conducting our study.

\subsection{Limitations}
One limitation of our work is that an attacker in the native attack scenario likely needs to get a Rowhammer enclave signed by a signing entity, \eg Intel or a BIOS vendor, to be able to launch the enclave.
While this sounds like a solid solution to prevent Rowhammer attacks through enclaves in practice, investigations on a very similar setting show that this is not the case~\cite{Chen2015}.
It is very well possible to slip malware into app stores~\cite{Chen2015}. 
Furthermore, most works on applications of SGX suggest that it can be used to keep the code and data secret from any third party~\cite{Schuster2015,Arnautov2016,Lee2017}.
Especially for secure cloud computation it is not plausible to run only signed enclaves, \ie a cloud provider will run non-signed user enclaves.
This would allow an attacker to run our attack as well.
Consequently, a different solution must be found to prevent Rowhammer attacks through SGX enclaves.

Although far more stealthy than spraying and grooming, memory waylaying is still observable by the OS.
The OS could prevent allocating too many page cache pages in a single process.
However, high memory requirements could also be perfectly reasonable, \eg trusted video processing~\cite{Lal2013}, operations on large encrypted database files~\cite{Schuster2015,Ohrimenko2016,Kerschbaum2017,Brekalo2016}.
Hence, it is unclear whether memory allocation patterns alone are enough to give away a Rowhammer attack.
There is no further interaction between the enclave and the non-enclave sides that could be monitored to detect the attack.
Finally, future software defenses may still prevent our attack, \eg by checking the integrity of binaries and terminating processes when an integrity check fails.

SGX enclaves should only be run if they are signed by Intel or a trusted partner.
If Intel or one of the trusted partners do not thoroughly review the code before singing it, our attack might slip through the signing process.
However, as this enclave signing process has not yet been deployed, it is unclear whether such a code review would  actually happen.
Perhaps more devastating is that fact that  users and businesses can deliberately run non-signed enclaves.
In fact, Microsoft already does this on the \emph{Azure confidential computing} cloud~\cite{AzureSGX}.
Hence, it is unclear whether a signing process would pose any limitation for our attack.

Currently, in our opcode flipping technique, the identification of target bit flip locations in binaries requires some manual work.
That is, manually defining a range where bit flips should be tested and manually selecting the groups of successful execution results.
While this is certainly feasible for a small number of binaries, fully automating this process would allow a complete analysis of the attack surface.
Similarly, compilers could generate code which guarantees that an attacker requires at least $N$ bit flips to successfully manipulate the control flow, \ie $N$ is a security parameter (\cf \cite{Barry2016compilation,Chen2017camfas}).
We consider this an interesting direction of future work, not only for research on Rowhammer attacks but also on fault attacks in general.

\subsection{Rowhammer mitigations in hardware}\label{sec:discussion_mitigations}
While it might be possible to design a practical software-based Rowhammer countermeasure, the results of our paper indicate that this is difficult, since not all variants of triggering the Rowhammer bug are known.
Furthermore, future Rowhammer defenses should also be designed with related fault attacks in mind~\cite{Kurmus2017,Karimi2015}.
We now discuss proposed and existing countermeasures implemented that require hardware modifications, but tackle the problem at its root.

ECC RAM can detect and correct 1-bit errors and, thus, deal with single bit flips caused by the Rowhammer attack.
Furthermore, IBM's Chipkill error correction~\cite{Chipkill} allows to successfully recover from 3-bit errors.
However, uncorrectable multi-bit flips can be exploitable~\cite{Aichinger2015hpec,Aichinger2015memcon,Lanteigne2016} or can result in a denial-of-service attack similar as described in~\cref{sec:dos} depending on how the OS responds to the error.
While only modern AMD Ryzen processors support ECC RAM in consumer hardware, Intel restricts its support to server CPUs, thus, making it unavailable in commodity systems.

While the LPDDR4~\cite{LPDDR4} implements TRR and MAC, \citeA{VanDerVeen2016CCSPresentation} still reported bit flips on a Google Pixel phone with \SI{4}{\giga B} LPDDR4 memory.
Doubling the refresh rate has been shown to be insufficient~\cite{Aweke2016,Kim2014} and a further increase would incur a too high performance penalty~\cite{Kim2014}.

\citeA{RAIM2012} introduced a redundant array of independent memory (RAIM) system as a feature of IBM's zEnterprise servers, which is basically the memory-equivalent for RAID systems for hard disks.
For an uncorrectable error, an attacker would have to induce multiple bit flips in different rows of different modules, making Rowhammer attacks infeasible.

\citeA{Kim2014} and \citeA{Kim2015} proposed to eliminate bit flips in hardware by probabilistically opening adjacent or non-adjacent rows, whenever a row is opened or closed.
As ongoing Rowhammer attacks open and close a certain row repeatedly, the vulnerable adjacent rows would be refreshed before bit flips occur.
Their approaches are possible solutions to mitigate Rowhammer attacks in future hardware.

\subsection{Design of SGX}
Intel SGX aims at protecting code from untrusted third parties.
Indeed, we see that it perfectly hides our attack from different defense mechanisms.
While this is intentional behavior and shows that SGX works, the question arises how to cope with harmful code within SGX enclaves, which eventually will happen in the wild.

A more discerning problem of SGX is that it halts the entire system, \eg a cloud system.
This is a powerful tool for attackers regardless of whether they run in the normal world or within an SGX enclave.
Taking down entire clouds, possibly in a coordinated and distributed way, poses a security risk.
Instead of halting the system, it would be less dangerous for the provider to only stop the running enclaves and return corresponding error codes to the host application.
A similar design change was also proposed by \citeA{Jang2017SGXBomb}.
\section{Conclusion}\label{sec:conclusion}
In this paper, we showed that even a combination of all state-of-the-art Rowhammer defenses does not prevent Rowhammer attacks.
Our novel attack and exploitation primitives systematically undermine the assumptions of all defenses.
With one-location hammering, we showed that previous assumptions on how the Rowhammer bug can be triggered are invalid and keeping only one DRAM row constantly open is sufficient to induce bit flips.
With a slow-down factor of only $3.3$, it is still on par with previous (now mitigated) techniques.
With opcode flipping, we bypass all memory layout-based defenses by flipping bits in a predictable and targeted way in the userspace \texttt{sudo} binary.
We present 29 bit offsets, each allowing an attacker to obtain root privileges in practice.
With memory waylaying, we present a reliable technique to replace conspicuous and unstable memory spraying and grooming techniques.
Coaxing the OS into relocating any binary page takes \SI{2.68}{\second} with our stealth-optimized variant, and only \SI{36.7}{\micro\second} with our speed-optimized variant.
Finally, we leveraged Intel SGX to hide the full privilege-escalation attack, making any inspection or detection of the attack infeasible.
Consequently, our attack evades all previously proposed countermeasures for commodity systems.

\section*{Acknowledgments}
We would like to thank our anonymous reviewers and Mark Seaborn for their valuable feedback as well as Thomas Schuster for help with some experiments.
This work has been supported by the Austrian Research Promotion Agency (FFG) via the K-project DeSSnet, which is funded in the context of COMET -- Competence Centers for Excellent Technologies by BMVIT, BMWFW, Styria and Carinthia.
This project has received funding from the European Union's Horizon 2020 research and innovation programme under European Research Council (ERC) grant agreement No 681402 and under grant agreement No 644052 (HECTOR).
Yuval Yarom performed part of this work as a visiting scholar at the University of Pennsylvania,
supported by an Endeavour Research Fellowship from the Australian Department of Education and Training.
Daniel Genkin was supported by NSF awards \#1514261 and \#1652259, financial assistance award 70NANB15H328 from the U.S. Department of Commerce, National Institute of Standards and Technology, the 2017-2018 Rothschild Postdoctoral Fellowship, and the Defense Advanced Research Project Agency (DARPA) under Contract \#FA8650-16-C-7622.

\bibliographystyle{IEEEtranSN}
\bibliography{bibliography/main}

\FloatBarrier

\appendix

\subsection{Bitflips in sudo}\label{app:bitflips_sudo}

\cref{tab:bitflips_sudo} lists exploitable bitflip offsets that modify opcodes of \texttt{sudoers.so} (Ubuntu 17.04, \texttt{sudo} version 1.8.19p1) yielding a skip of the privilege check and, thus, elevating an unprivileged process to root privileges. 

\begin{table*}[t]
  \centering
  \caption{Exploitable bitflip offsets in \texttt{sudoers.so}.}\label{tab:bitflips_sudo}
  \begin{tabular}{rccll}
    \toprule
    \thead{\#} & \thead{Binary offset} & \thead{Bitflip offset} & \thead{Original} & \thead{Flipped} \\
    \midrule
      1  & \texttt{0x8c1c} & 4 & \texttt{lea rdi, aUser\_is\_exempt}         & \texttt{lea rbp, aUser\_is\_exempt} \\
      2  & \texttt{0x8c32} & 3 & \texttt{mov eax, ebp}                       & \texttt{mov eax, esp} \\
      3  & \texttt{0x8d4e} & 0 & \texttt{lea rax, off\_250860}               & \texttt{lea rax, off\_250860+1} \\
      4  & \texttt{0x8d4f} & 0 & \texttt{lea rax, off\_250860}               & \texttt{lea rax, unk\_250760} \\
      5  & \texttt{0x8d59} & 0 & \texttt{mov eax, [rax+2C8h]}                & \texttt{mov eax, [rax+2C9h]} \\
      6  & \texttt{0x8d59} & 1 & \texttt{mov eax, [rax+2C8h]}                & \texttt{mov eax, [rax+2CAh]} \\
      7  & \texttt{0x8d59} & 2 & \texttt{mov eax, [rax+2C8h]}                & \texttt{mov eax, [rax+2CCh]} \\
      8  & \texttt{0x8d59} & 3 & \texttt{mov eax, [rax+2C8h]}                & \texttt{mov eax, [rax+2C0h]} \\
      9  & \texttt{0x8d59} & 6 & \texttt{mov eax, [rax+2C8h]}                & \texttt{mov eax, [rax+288h]} \\
      10 & \texttt{0x8d5a} & 5 & \texttt{mov eax, [rax+2C8h]}                & \texttt{mov eax, [rax+22C8h]} \\
      11 & \texttt{0x8d5d} & 7 & \texttt{test eax, eax}                      & \texttt{add eax, 485775C0h} \\
      12 & \texttt{0x8d5e} & 0 & \texttt{test eax, eax}                      & \texttt{test ecx, eax} \\
      13 & \texttt{0x8d5f} & 0 & \texttt{jnz short check\_user\_is\_exempt } & \texttt{jz short check\_user\_is\_exempt} \\
      14 & \texttt{0x8dbd} & 3 & \texttt{test al, al}                        & \texttt{mov eax, es} \\
      15 & \texttt{0x8dbd} & 7 & \texttt{test al, al}                        & \texttt{add al, 0C0h} \\
      16 & \texttt{0x8dbf} & 0 & \texttt{jnz short near ptr unk\_8D61}       & \texttt{jz short near ptr unk\_8D61} \\
      17 & \texttt{0x8dbf} & 3 & \texttt{jnz short near ptr unk\_8D61}       & \texttt{jge short near ptr unk\_8D61} \\
      18 & \texttt{0x8dc4} & 3 & \texttt{lea rbp, qword\_252700}             & \texttt{lea rbp, algn\_2526F8} \\
      19 & \texttt{0x8dc5} & 1 & \texttt{lea rbp, qword\_252700}             & \texttt{lea rbp, dword\_252900} \\
      20 & \texttt{0x8dc5} & 2 & \texttt{lea rbp, qword\_252700}             & \texttt{lea rbp, \_\_imp\_fflush} \\
      21 & \texttt{0x8dc9} & 3 & \texttt{mov eax, [rbp+0F0h]}                & \texttt{mov ecx, [rbp+0F0h]} \\
      22 & \texttt{0x8dc9} & 4 & \texttt{mov eax, [rbp+0F0h]}                & \texttt{mov edx, [rbp+0F0h]} \\
      23 & \texttt{0x8dca} & 7 & \texttt{mov eax, [rbp+0F0h]}                & \texttt{mov eax, [rbp+70h]} \\
      24 & \texttt{0x8dcb} & 3 & \texttt{mov eax, [rbp+0F0h]}                & \texttt{mov eax, [rbp+8F0h]} \\
      25 & \texttt{0x8dcf} & 0 & \texttt{test eax, eax}                      & \texttt{test ecx, eax} \\
      26 & \texttt{0x8dcf} & 3 & \texttt{test eax, eax}                      & \texttt{test eax, ecx} \\
      27 & \texttt{0x8dd0} & 2 & \texttt{jnz loc\_8FB0}                      & \texttt{or eax, [rbp+1DAh]} \\
      28 & \texttt{0x8dd1} & 0 & \texttt{jnz loc\_8FB0}                      & \texttt{jz loc\_8FB0} \\
      29 & \texttt{0x8e23} & 6 & \texttt{jz loc\_8FE8}                       & \texttt{jz near ptr algn\_8FA7+1} \\
    \bottomrule
  \end{tabular}
\end{table*}

\subsection{Computing the Optimal Runtime of our Attack}\label{app:attack_runtime}

The runtime of our attack is computed as
\[\frac{P \cdot (W + n \cdot 0.05)}{2^{12} \cdot n}+\frac{n \cdot 2^{16}}{F \cdot E} + \frac{120\cdot P}{2^{30}}\]
seconds, where $P$ is the amount of physical memory installed in the system, $W$ is the amount of time one waylaying relocation takes, $F$ is the flip rate (\ie bit flips per second), and $E$ is the number of exploitable bit offsets within a \SI{4}{\kilo B} page (which depends on the target binary).
$n \in \mathbb{N}$ is the optimization parameter, the number of bit flips to find in the templating phase, influencing the runtime of the templating phase and the waylaying phase.
$0.05$ seconds is the time the prefetch address-translation oracle consumes for one test.
$120$ seconds is the amount of time the prefetch side-channel attack consumes to translate a virtual to a physical address per gigabyte ($2^{30}$ bytes) of system memory.
The $2^{16}$ represent the $2^{15}$ bit offsets of a \SI{4}{\kilo B} page ($2^{12}$ bytes) which can flip in both directions each.

On our test system we have $P=12$ gigabytes, $W=2.68$ seconds for memory waylaying, $F=0.67$, and $E=29$.
With these values we compute the runtime as
\[\frac{3 \cdot 2^{20} \cdot (2.68 + n \cdot 0.05)}{n}+n \cdot 3373.3 + 24\,\textnormal{m}\]
seconds.
The minimum of this function is reached at $n=50$.

\begin{figureA}[t]{bitflip-runtime}[Expected total runtime (templating and waylaying) until
  the attacker has the target page at the target physical location.]
 \begin{tikzpicture}
\begin{axis}[
mlineplot,
style={font=\footnotesize},
xlabel={Exploitable bitflips},
ylabel={Runtime [hours]},
width=1.0\hsize,
xmin=0,
ytick={0,100,...,500},
xtick={0,10,...,100},
xmax=100,
ymax=570,
ymin=0,
height=4cm,
legend columns=3,
legend style={font=\tiny,at={(0.545,1.002)}, anchor=north}]
]
\addplot+[blue, mark=o,mark phase=2,mark repeat={6}, restrict x to domain=1:100] table[x=Bitflips,y=zs,col sep=comma]{data/bitflip_runtime.csv};
\addlegendentry{One-location}
\addplot+[red, mark=x,mark phase=1,mark repeat={6}, restrict x to domain=1:100] table[x=Bitflips,y=ds,col sep=comma] {data/bitflip_runtime.csv};
\addlegendentry{Double-sided}
\addplot+[black, mark=diamond,mark phase=3,mark repeat={6},restrict x to domain=1:100] table[x=Bitflips,y=ss,col sep=comma] {data/bitflip_runtime.csv};
\addlegendentry{Single-sided}
\addplot+[blue, densely dashed, thick, mark=o,mark phase=2,mark repeat={6}, restrict x to domain=1:100] table[x=Bitflips,y=zs-mc,col sep=comma]{data/bitflip_runtime.csv};
\addlegendentry{One-location (MC)}
\addplot+[red, densely dashed, thick, mark=x,mark options={fill=red},mark phase=1,mark repeat={6}, restrict x to domain=1:100] table[x=Bitflips,y=ds-mc,col sep=comma] {data/bitflip_runtime.csv};
\addlegendentry{Double-sided (MC)}
\addplot+[black, densely dashed, thick, mark=diamond,mark phase=3,mark repeat={6},densely dotted,restrict x to domain=1:100] table[x=Bitflips,y=ss-mc,col sep=comma] {data/bitflip_runtime.csv};
\addlegendentry{Single-sided (MC)}
\end{axis}
\end{tikzpicture}
\end{figureA}

\cref{fig:bitflip-runtime} shows the expected total runtime of the templating phase, and memory waylaying and chasing, depending on which hammering technique is used and how many bit offsets are exploitable.

\subsection{Memory Basics, Policies, and their Influence on One-Location Hammering}\label{app:memory_controller_policies}
DRAM is organized in multiple banks, \eg for a dual-channel dual-rank configuration 32 banks on DDR3 and 64 banks on DDR4.
Each bank consists of an array of rows of \SI{8}{\kilo B} each.
Thus, the number of rows is typically in the range of $2^{14}$ to $2^{16}$.
Since the DRAM cells lose their charge over time, the DDR standard defines that every row must be refreshed once per \SI{64}{\micro \second}.
When accessing a memory location, the corresponding row is opened, \ie copied into an internal array called the \emph{row buffer}.
Closing a row copies the data from the row buffer back into the actual DRAM cells.

Before a row can be opened, the bank has to be precharged.
Consequently, when accessing a memory location in the currently opened row, \ie a row hit, the latency is comparably low.
Accessing a memory location in a different row, \ie a row conflict, incurs first closing the DRAM row, then precharging the bank, and finally opening the new row, copying the data into the row buffer.
The latency in this case is significantly higher, \eg \SI{200}{\percent} of the latency of a row hit.

The memory controller can optimize the memory performance by cleverly deciding when to close a row preemptively.
The two most basic memory controller policies are ``open page'' and ``closed page''.
An open-page policy keeps the recently accessed row open and buffered.
This is beneficial for memory access latency, power consumption, and bank utilization when the number of memory accesses is low~\cite{Kaseridis2011}.
However, when the number of memory accesses increases the situation is more complex.
A closed-page policy can then achieve a better system performance, since the row is immediately closed and the bank is precharged and ready to open a new row~\cite{Kaseridis2011}.

With modern processors having huge caches and complex algorithms for spatial and temporal prefetching, the probability that further memory accesses go to the same row decreases.
Consequently, more complex memory controller policies have been proposed and are implemented in modern processors~\cite{Kaseridis2011}.
David~\etal\cite{David2011} noted that closed-page policies perform especially better on multi-core systems and hence they assumed that these are implemented in current processor architectures.
Intel also holds patents for dynamically adjusting memory controller policies~\cite{Kahn2004}.
A closed-page policy, but also other policies which preemptively close rows, would allow one-location hammering.

Besides these memory controller policies, the memory controller can also reorder and combine memory accesses~\cite{Rotithor2006}.
Since the Rowhammer bug is related to the number of row activations~\cite{Kim2014}, a lower number of activations due to reordering and combining also reduces the probability of bit flips.
In one-location hammering most of the accesses can be expected to be reordered and combined to reduce the overall number of row activations, leading to a lower number of bit flips than with other hammering techniques.

\end{document}